\documentclass[9pt]{article}
\usepackage{amssymb, amsmath}
\usepackage[dvipdfm]{graphicx}
\begin{document}

\makeatletter
%%%%%------- preprint style ----%%%%%%%%%%%%%%%%%%%%
\@addtoreset{equation}{section}
\def\theequation{\thesection.\arabic{equation}}
\def\@maketitle{\newpage
 \null
 {\normalsize \tt \begin{flushright} 
  \begin{tabular}[t]{l} \@date  
  \end{tabular}
 \end{flushright}}
 \begin{center} 
 \vskip 2em
 {\LARGE \@title \par} \vskip 1.5em {\large \lineskip .5em \begin{tabular}[t]{c}\@author 
 \end{tabular}\par} 
 \end{center}
 \par
 \vskip 1.5em} 
\makeatother
%%%%%%%%%%%%%%%%%%%%%%%%%
\topmargin=-1cm
\oddsidemargin=1.5cm
\evensidemargin=-.0cm
\textwidth=15.5cm
\textheight=22cm
%\renewcommand{\baselinestretch}{1.5}
%%%%%%%%%%%%%%%%%%%%%%%%
\setlength{\baselineskip}{16pt}
\title{Quantization of a Scalar Field in Two Poincar\'e Patches of Anti-de Sitter Space and AdS/CFT}
\author{
Ippei~{\sc Fujisawa}\thanks{ifujisawa@particle.sci.hokudai.ac.jp} and
Ryuichi~{\sc Nakayama}\thanks{nakayama@particle.sci.hokudai.ac.jp}
       \\[1cm]
{\small
    Division of Physics, Graduate School of Science,} \\
{\small
           Hokkaido University, Sapporo 060-0810, Japan}
}
\date{
EPHOU-14-004  \\
%March  2014 @@
}
% 
%\begin{titlepage}
% 
\maketitle

\begin{abstract} 
Two sets of modes of a massive free scalar field are quantized in a pair of Poincar\'e patches of Lorentzian anti-de Sitter (AdS) space, AdS$_{d+1}$ ($d \geq 2$). It is shown that in Poincar\'e coordinates $(r,t,\vec{x})$, the two boundaries at $r=\pm \infty$ are connected. When the scalar mass $m$ satisfies a condition $0 < \nu=\sqrt{(d^2/4)+(m\ell)^2} <1$, there exist two sets of mode solutions to Klein-Gordon equation, with distinct fall-off behaviors at the boundary. 
By using the fact that the boundaries at $r=\pm \infty$ are connected, a conserved Klein-Gordon norm can be defined for these two sets of scalar modes, and these modes are canonically quantized. Energy is also conserved.  A prescription within the approximation of semi-classical gravity is presented for computing two- and three-point functions of the operators in the boundary CFT, which correspond to the two fall-off behaviours  of scalar field solutions.

\end{abstract}

%\vspace*{2cm}


%\end{titlepage}
\newpage
\setlength{\baselineskip}{18pt}

%%%%%%%%%%%%%%%%%%%%%%%%%%%%%%%%%%%%%%%%%%%%%%%%%%%%%%%%%%%%%%%%%%%%%
\newcommand{\bm}[1]{\mbox{\boldmath $#1$}}
%%%%%%%%%%%%%%%%%%%%%%%%%%%%%%%%%%%%%%%%%%%%%%%%%%%%%%%%%%%%%%%%%%%%

\section{Introduction}
\hspace{5mm}
Quantization of scalar fields propagating in anti-de Sitter space was attempted in the past. \cite{AIS}\cite{BF}\cite{BKL} In \cite{AIS} the problem of a time-like boundary at space-like infinity, through which data can propagate, is studied in a massless case by conformally mapping the spacetime in the global coordinates into upper hemisphere of Einstein static universe (ESU). By using the fact that AdS space is mapped to a half of ESU, it was shown that there are two sets of mode functions, which are characterized by different boundary conditions and are orthonormal and form a complete set of basis by themselves, separately. It was concluded that only one of the two sets of mode functions can be quantized. In \cite{BF} this procedure is more elaborated and extended to massive scalars. In \cite{BKL} mode functions for scalar fields in AdS space in both  Poincar\'e coordinates and global coordinates are obtained. Group theoretic analysis was performed in\cite{Fronsdal}.

On the other hand, AdS/CFT duality was discovered in\cite{ads1} and its  precise definition has been developed.\cite{ads2}\cite{ads3}\cite{ads4}\cite{ads5} To the string compactification on AdS$_{d+1} \times {\cal M}_n$, there corresponds a conformal field theory (CFT) living on a space conformal to the d-dimensional boundary of the AdS.  To each field $\Phi$ in the bulk there corresponds a local operator in the CFT. By fixing the boundary value of $\Phi$ and computing the effective action of the bulk theory, this effective action yields the generating functional of the operators in conformal field theory with the boundary value acting as the source function. In the semiclassical supergravity limit, one can compute the effective action by solving the classical equation of motion and just substituting the solution into the action.

In the case of a free scalar field $\phi$ of mass $m$, it falls off like $\phi \sim r^{-(d-\Delta_+)} \, \phi_0+r^{-(d-\Delta_-)} \, \phi_1$ near the spacelike boundary $r \rightarrow \infty$. Here $\Delta_{\pm}=\frac{d}{2}\pm \nu$ and $\nu =\sqrt{(d^2/4)+m^2}$. When $\nu >1$, only $\phi_0$  acts as a source for an operator ${\it O}_+$ with a scaling dimension $\Delta_+$ in CFT. When $0 < \nu < 1$, it is argued that either of the two operators ${\it O}_+$ and ${\it O}_-$ with scaling dimensions $\Delta_+$ and $\Delta_-$ can be considered in CFT.  To compute two-point functions of ${\it O}_+$ one should take $\phi_0$ as a source function and functionally differentiate the effective action with respect to $\phi_0$.\cite{ads3} To compute two-point functions of ${\it O}_-$, however,  one needs to Legendre transform the effective function with respect to $\phi_0$ to obtain a generating functional.\cite{ads4} This restriction of the holographic correspondence is argued to be related to the above peculiarity of the scalar field quantization in AdS space. 

Meanwhile, in the context of  AdS/CFT for 3d higher-spin gravity coupled to matter fields,  it was found\cite{FNN3} that we can compute semi-classically two-point functions of two sets of single-trace operators in boundary CFT by introducing only one set of matter fields $B$ and $C$.  This motivates us to study whether we can quantize a scalar field in AdS space while keeping both two sets of scalar modes.  

One of the purposes of this paper is to show that these two sets of scalar modes in AdS space can be quantized altogether by considering a coordinate system which is obtained by patching together a pair of Poincar\'e coordinates with radial coordinate $r >0$ and $r<0$, respectively, along the horizon ($r=0$). The AdS space can be divided into two Poincar\'e patches. The boundary of AdS space is also divided into two. Usually, a scalar field is quantized only in one of the two Poincar\'e patches. 
%However, Poincar\'e patch is not simply one of the two sides of the horizon of a hyperboloid which represents AdS. One Poincar\'e patch is composed of some regions of both sides of the horizon on the hyperboloid.  
In connection with AdS/CFT correspondence, however, conformal symmetry of boundary CFT has an origin in the isometry of AdS space. Although the metric in a pair of Poincar\'e coordinates is invariant under special conformal transformations, points in the two Poincar\'e patches are exchanged and a single Poincar\'e patch is not invariant. Hence it is not appropriate to restrict analysis of a field theory in AdS space to just within a single patch.\footnote{ In \cite{Gibbons} a quotient space AdS$_{d+1}$/$J$, where $J$ is an antipodal map $X_{\mu} \rightarrow -X_{\mu}$, is considered. This space is invariant under the isometry of AdS$_{d+1}$.}$^,$\footnote{EAdS space is one piece of the two disconnected hyperbolic spaces, and this single piece has the full conformal symmetry. This is in sharp contrast to the Lorentzian case. } 

In this paper, it is shown that the two patches can be joined together by matching the fluxes of a scalar field across the horizon and two boundaries, and that the united coordinate system admits two sets of scalar mode functions.  The fluxes across the horizon vanishes, while those across the boundaries do not. These fluxes across the two boundaries, however, cancel out with each other. It is shown that in Poincar\'e coordinates for AdS$_{d+1}$ ($d\geq 2$), the two boundaries $r=\pm \infty$ are connected.   Hence the cancellation of the fluxes occurs on the connected boundaries of the hyperboloid. As a result, Klein-Gordon norm (\ref{KGnorm}) is conserved. 
It is also shown that energy is conserved.

After canonical quantization of the scalar field,  Wightman function for a scalar field in AdS space is computed by performing explicit integrations.\footnote{Wightman function for a scalar field in AdS space was computed previously by solving a differential equation with respect to an AdS-invariant distance and matching its singularity with that of flat space.\cite{BFDG}} 
An allowed form of boundary conditions for a scalar field on the two boundaries is also identified. 
An interesting issue of AdS/CFT is the prescription for semi-classically computing two-point functions of ${\it O}_+$ and ${\it O}_-$ for a scalar field theory with a mass in the range $-d^2/4 < m^2<1-d^2/4$. To present  this prescription is the second aim of this paper. It turns out that the (renormalized) action integral (in Euclidean AdS (EAdS) space) is given by a sum of bulk action and  boundary terms. 
\begin{multline}
I=  \int_{-\infty}^{\infty} dr \, \int d^d y \sqrt{g} \, \Big(\frac{1}{2}  \,  g^{\mu\nu} \,\partial_{\mu} \phi\, \partial_{\nu}\phi+\frac{1}{2} \, m^2\phi^2\Big) 
+ \lim_{r \rightarrow + \infty}   \int_{r \ \text{fixed}} d^d\vec{y} \, \sqrt{\gamma} \, \frac{1}{2}  \,  \Delta_- \, \phi^2  \\
-\lim_{r \rightarrow - \infty}   \int_{r \ \text{fixed}} d^d\vec{y} \, \sqrt{\gamma} \, \frac{1}{2} \,  \Delta_- \, \phi^2  
-\lim_{r \rightarrow - \infty}   \int_{r \ \text{fixed}} d^d\vec{y} \, \sqrt{\gamma} \, \phi \, r \, \partial_r \, \phi .  
\end{multline}
Here $r$ is the radial coordinate which takes the value in the range $-\infty <r <\infty$. $r=0$ is the horizon and $r=\pm \infty$ are the two boundaries. Two  Poincar\'e patches are also introduced in the EAdS space corresponding to the Lorentzian version. The metric is given by $ds^2=dr^2/r^2+r^2 \, d\vec{y}^2=g_{\mu\nu}dy^{\mu}dy^{\nu}$ and $\gamma_{ij}$ is an induced metric on the boundaries and $\sqrt{\gamma}=|r|^d$. 
The $\phi^2$, $\phi \, \partial_r \, \phi$ terms on the boundaries are counterterms to cancel out the divergences which appear in calculation of the two-point functions. Two boundary values $\phi_+$, $\phi_-$ of a scalar field will be used as source functions for the two-point functions in boundary CFT. Legendre transformation is not required. Calculation of  three-point functions  with our formalism is also outlined. 

This paper is organized as follows. In sec.2 a global coordinates and Poincar\'e coordinates of AdS space are reviewed and peculiar properties of Lorentzian AdS space in Poincar\'e coordinates are discussed. A prescription for patching together two Poincar\'e charts is explained.    In sec.3 Klein-Gordon (KG) equation will be solved in each Poincar\'e patch, and two kinds of mode functions in a pair of Poincar\'e patches are determined in such a way that KG norm is conserved. It is checked that the fluxes through the horizon vanish, and the fluxes at the boundaries cancel out. Conservation of energy is also shown.  
In sec.4 a scalar field operator is expanded into these modes, and canonical commutation relations are applied. 
In sec.5, Wightman function of a scalar field is computed explicitly. AdS/CFT correspondence for two-point functions will be studied in sec.6. Due to the properties of the mode functions obtained in sec.3, the solutions to the equation of motion on the pair of Poincar\'e patches have a peculiar parity property with respect to the radial coordinate $r$, which is  modified by a parameter $S$. This fact allows us to write down a general solution $\phi$ in terms of two boundary values $\phi_+$, $\phi_-$ of the scalar field.  By assuming some form of boundary actions on the two boundaries,  substituting the solution into the action, and adjusting the coefficients of the boundary terms to eliminate divergences as $|r| \rightarrow \infty$, we get a suitable generating functional of two-point functions. A prescription which makes both two point functions $\langle {\it O}_+ {\it O}_+ \rangle$ and $\langle {\it O}_- {\it O}_- \rangle$ positive is proposed. In sec.7 a prescription for computing three-point functions in a bulk $\phi^3$ theory is mentioned.  Sec.8 is devoted to a summary and discussions. In appendix A, an explicit calculation of Wightman function is presented. In Appendix B a method for calculating  integrals of products of the bulk-boundary propagators $K_{\Delta_{\pm}}$ is outlined. 

\section{AdS Spacetime}
\hspace{5mm}
\subsection{Definition}
\hspace{5mm}
A $d+1$-dimensional AdS spacetime  AdS$_{d+1}$ is defined by a constant negative curvature hyperboloid 
\begin{equation}
X \cdot X \equiv -X_0^2-X_{d+1}^2+\sum_{i=1}^{d}\, X_i^2=-\ell^2  \label{hyperboloideq}
\end{equation}
embedded in pseudo-Minkowski space $\mathbb{E}^{d,2}$. Here $\ell$ is an AdS radius. Line element in $\mathbb{E}^{d,2}$ induces a one on this hyperboloid.  
\begin{equation}
ds^2 = -dX_0^2-dX_{d+1}^2+\sum_{i=1}^d dX_i^2
\end{equation}
There are several coordinate systems, and the global coordinates and the Poincar\'e ones are among them.\footnote{For review see for example \cite{BKL}, \cite{ads5}.} 

Global coordinates are defined by 
\begin{eqnarray}
X_0 &=& \ell \, \sec \rho \, \cos \tau, \nonumber \\
X_i &=& \ell \, \tan \rho \ \Omega_i, \ (i=1, \dots, d) \nonumber \\
X_{d+1} &=& \ell \, \sec \rho \, \sin \tau, \label{globalcoord}
\end{eqnarray}
where radial coordinate $\rho$ and time $\tau$ take values in ranges $0 \leq \rho < \pi/2$ and $-\pi < \tau \leq \pi$, and $\rho=\pi/2$ is a boundary. Spherical coordinates $\Omega_i$ satisfy  $-1 \leq \Omega_i \leq 1$ and $\sum_i \Omega_i^2=1$. The line element is given by 
\begin{equation}
ds^2 = \ell^2 \, \sec^2 \rho \, \Big(d\rho^2-d\tau^2+\sin^2 \rho \, \sum_{i=1}^d \, d\Omega_i^2\Big).
\end{equation}
To avoid time-like closed loops, one unwraps $\tau$ to have range $-\infty < \tau <\infty$, and works with a universal covering space, CAdS$_{d+1}$. 

Poincar\'e coordinates are defined by 
\begin{eqnarray}
X_0 &=& \frac{1}{2} z \, \Big(1+\frac{1}{z^2} \, (\ell^2-t^2+\vec{x}^2)\Big), \nonumber \\
X_d &=& \frac{1}{2} z\, \Big(1+\frac{1 }{z^2} \, (-\ell^2-t^2+\vec{x}^2)\Big), \nonumber \\
X_i &=& \frac{\ell}{z} \, x^i, \nonumber \\
X_{d+1} &=& \frac{\ell}{z}  \, t. \label{Poincare}  
\end{eqnarray}
Here $t$ and $x^i$ range between $-\infty$ and $\infty$, and radial coordinate $z$ ranges over $0 \leq z <\infty$. The line element is now given by 
\begin{equation}
ds^2 = \frac{\ell^2}{z^2} \, (dz^2-dt^2+d\vec{x}^2). \label{lineP}
\end{equation}
The boundary is at $z=0$. There is also a Killing horizon at $z=\infty$. The time-like Killing vector becomes null at this horizon. 

Poincar\'e coordinates cover only half of AdS$_{d+1}$, since $X_0-X_d=1/z>0$. The remaining half is covered by 
coordinates (\ref{Poincare}) with $-\infty <z \leq 0$. 
%To cover CAdS$_{d+1}$, one needs to consider an infinite sequence of pairs of Poincar\'e patches and put each pair on top of another. 
Usually, when AdS spacetime is studied in Poincar\'e patch, only a single patch is considered. However, as is explained below, it is necessary to consider a pair of Poincar\'e patches. 

\subsection{Two Poincar\'e Patches}
\hspace{5mm}

\begin{figure}[htbp]
\begin{minipage}{0.5\hsize}
\begin{center}
\includegraphics[scale=.3]{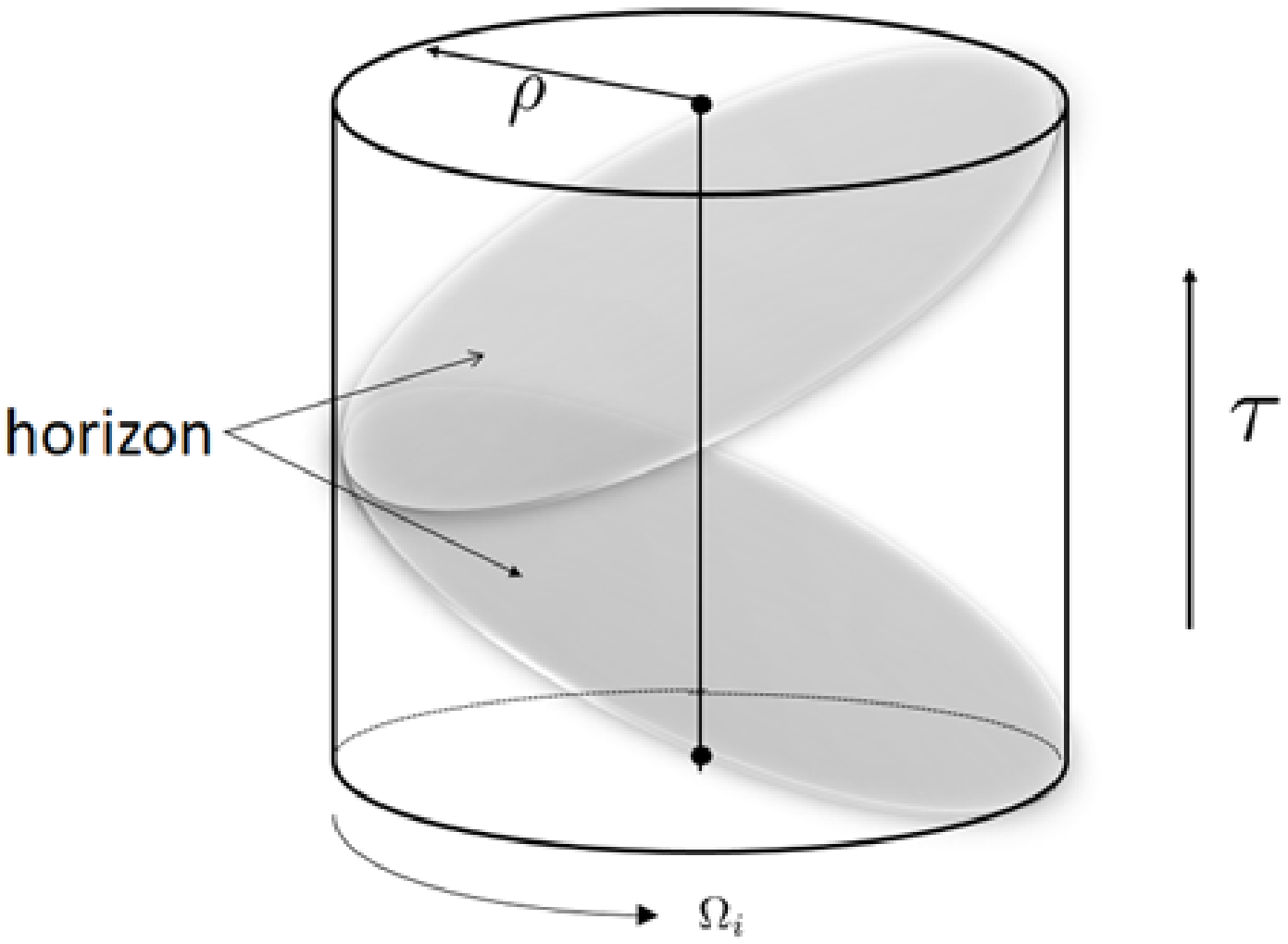} 
\end{center}
\caption{ The horizon in AdS is obtained by making two diagonal cuts through the cylinder. } \label{fig1}
\end{minipage}
\begin{minipage}{0.5\hsize}
\begin{center}
\includegraphics[scale=.3]{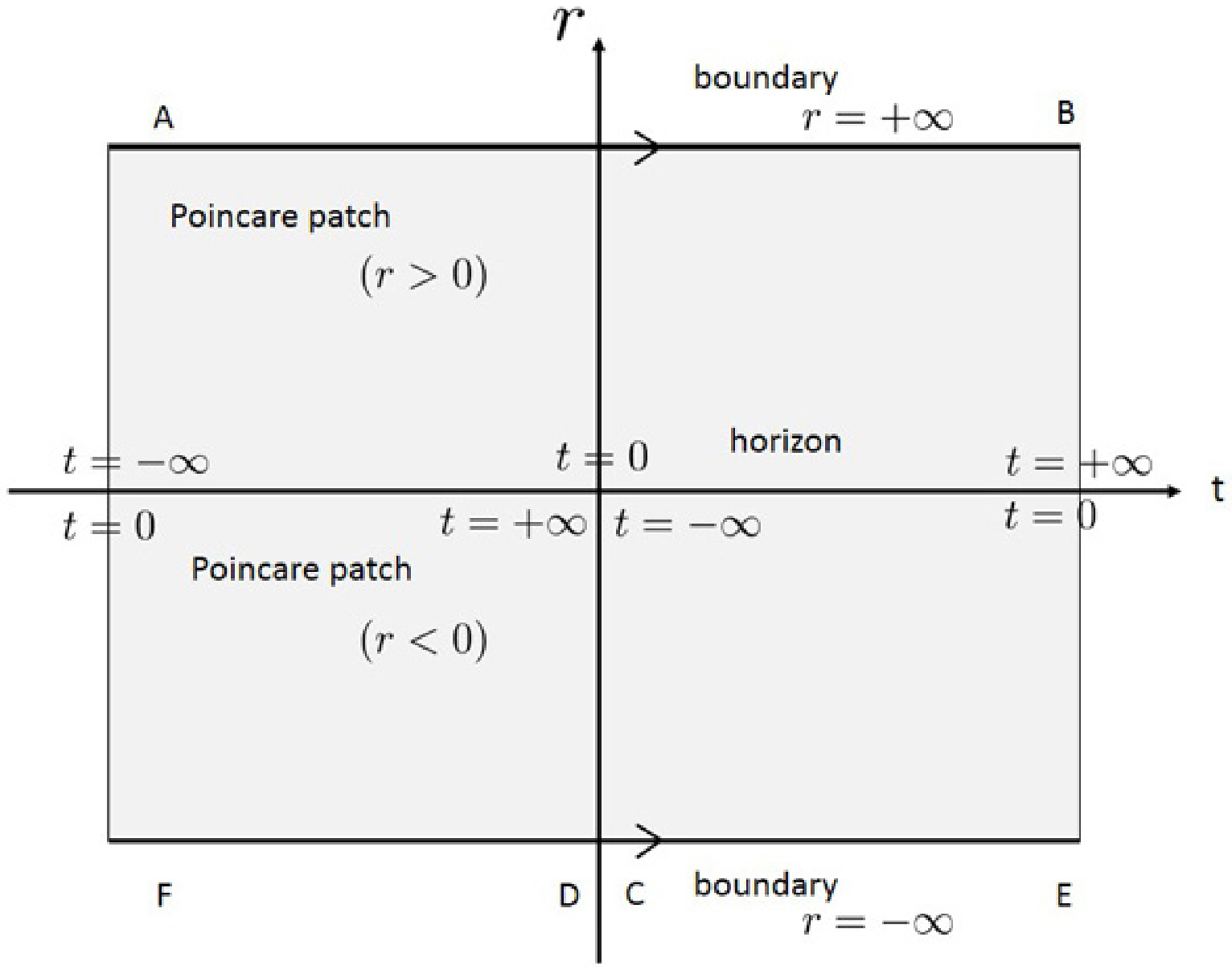} 
\end{center}
\caption{AdS space is constructed by gluing two Poincar\'e patches at the horizon. The time flows on the two Poincar\'e patches near the horizon are shifted with respect to each other. Only the $\vec{x}=\vec{0}$ section is displayed.}  \label{fig2}
\end{minipage}
\end{figure}

AdS space can be illustrated as an interior of a cylinder as in Figure \ref{fig1}. The boundary of AdS is identified with the boundary of the cylinder. The horizons in AdS are obtained by making two diagonal cuts through the cylinder. The cuts divide AdS into two regions, each of which is covered by each of a pair of Poincar\'e coordinates. 
By using a pair of Poincar\'e coordinates, a single cover of AdS space is obtained. A simplified view (with only $\vec{x}=\vec{0}$ section) is given in Figure \ref{fig2}. Here a new radial coordinate $r=\frac{1}{z}$ is introduced. This ranges over $-\infty <r<\infty$. The line element (\ref{lineP}) is rewritten as 
\begin{equation}
ds^2= \ell^2 \, \Big(r^{-2} \, dr^2+r^2(-dt^2+d\vec{x}^2)\Big) \equiv g_{\mu\nu} \, dx^{\mu}dx^{\nu}. \label{AdSmetric}
\end{equation}
The boundaries are at $r=\pm \infty$ and the horizon is at $r=0$. The metric (\ref{AdSmetric}) degenerates at the horizon $r=0$, but there is no singularity in the curvature tensor $R_{\mu\nu\lambda\rho}= \ell^{-2} \, (g_{\mu\lambda}g_{\nu\rho}-g_{\mu\rho}g_{\lambda\nu})$. 
Although the two boundaries in Figure \ref{fig2} are separated far apart, as will be shown below, points on some hypersurface of the boundaries must be identified. The conformal boundary of AdS$_{d+1}$ is a two-fold cover of conformally compactified Minkowski spacetime $\mathbb{E}^{d-1,1}$: $\partial (\text{AdS}_{d+1})= S^{d-1} \times S^1$ as in Figure \ref{fig1}. And that of the universal cover is Einstein static universe: $\partial (\text{CAdS}_{d+1}) = ESU_{d}$. 

\begin{figure}[htpb]
\begin{minipage}{0.5\hsize}
\begin{center}
\includegraphics[scale=0.3]{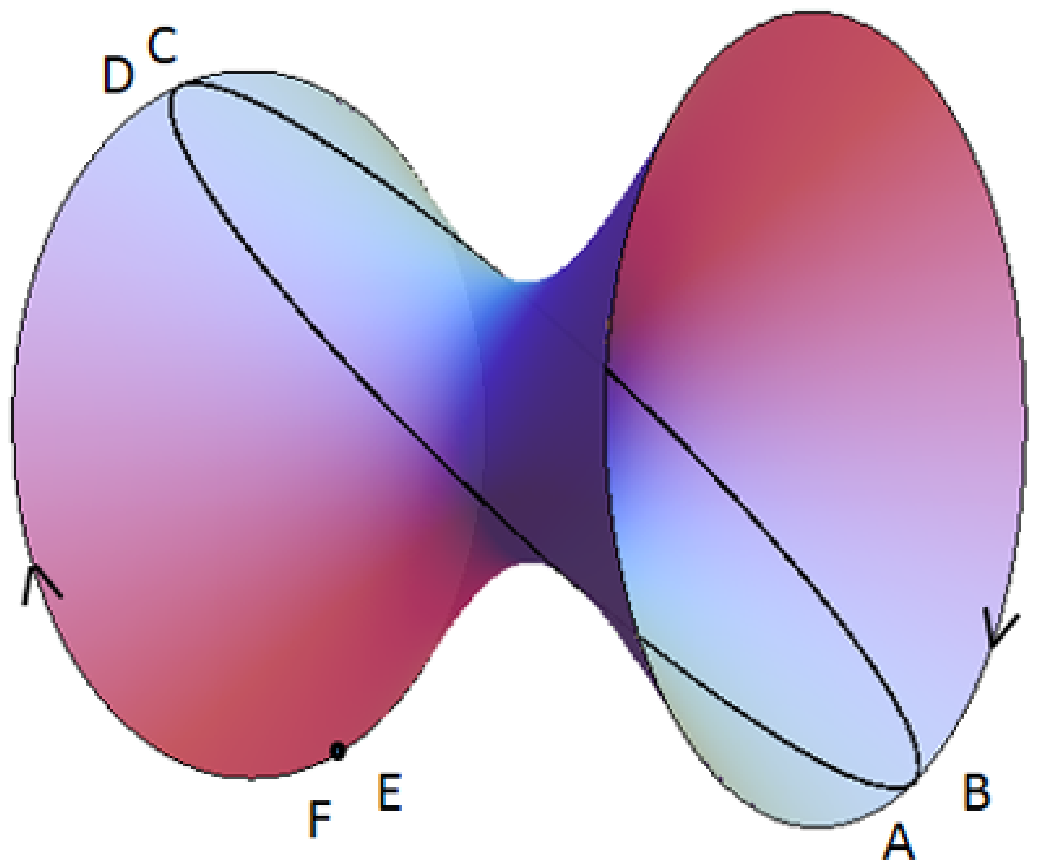} 
\end{center}
\caption{AdS space is represented as a hyperboloid. The oblique oval curve on the hyperboloid is the horizon. Points $A, \ldots ,F$ coorespond to those in Figure \ref{fig2}.} \label{fig3}
\end{minipage}
\begin{minipage}{0.5\hsize}
\begin{center}
\includegraphics[scale=0.6]{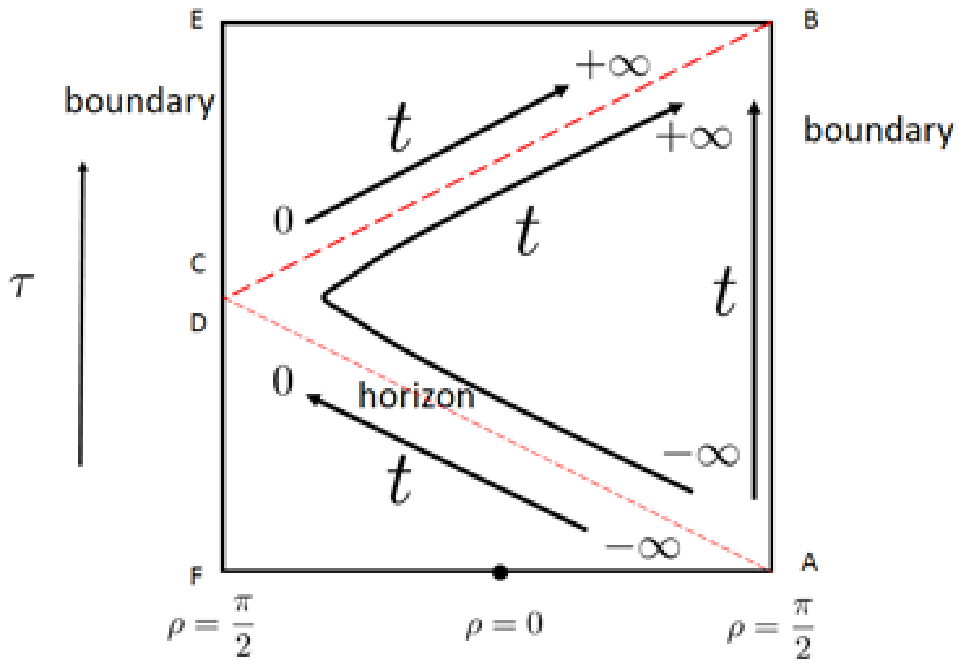} 
\end{center}
\caption{Penrose diagram of AdS; The flows of time $t$ are displayed. Points $A, \ldots ,F$ coorespond to those in Figure \ref{fig2}.}  \label{fig4}
\end{minipage}
\end{figure}

Furthermore,  we need to take into account the flows of time $t$.  Let us look at Figure \ref{fig4}. The left Poincar\'e patch in Figure \ref{fig4} is also a single region due to periodicity in $\tau$. The flows of time $t$ are displayed.  These flows are consistent with (\ref{Poincare}). The flows on the two Poincar\'e patches near the horizon are shifted with respect to each other by infinity, but we glue together the corresponding edges of the two Poincar\'e patches directly along the horizon. The resulting time coordinate is the one shown in Figure \ref{fig2}.

In general, time variables in two different patches separated by a horizon do not need to coincide. In the next section, it will be shown that the fluxes of a scalar field across the horizon from each Poincar\'e patch vanish. Hence even if the time coordinates in the upper and lower patches are different, the fluxes are matched on both sides of the horizon.

\subsection{Conformal Symmetry of Poincar\'e Patch}
\hspace{5mm}
Importance of introducing a pair of Poincar\'e patches is understood by the following observation. 
A single set of Poincar\'e coordinates do not preserve the full isometry of AdS$_{d+1}$ space, $SO(2,d)$, but only its subgroup $ISO(1,d-1) \times SO(1,1)$ (Poincar\'e and dilatation symmetries). However, by introducing two Poincar\'e charts, a special conformal transformation, 
\begin{eqnarray}
t  & \rightarrow & t' = \frac{ t+(x^2+r^{-2}) \, a^{0}   }
{1+a^2(x^2+r^{-2})+2a \cdot x}, \\
\vec{x}  & \rightarrow & \vec{x}\,' = \frac{   \vec{x}+(x^2+r^{-2}) \, \vec{a}   }
{1+a^2(x^2+r^{-2})+2a \cdot x}, \\
r & \rightarrow & r' =r \, \big( 1+2 a \cdot x+a^2 (x^2+r^{-2})\big),  \label{scr}
\end{eqnarray}
also becomes a symmetry transformation of (\ref{AdSmetric}), and full conformal symmetry is realized. $a^i=(a^0,\vec{a})$ is a constant vector. ($x^2\equiv-t^2+\vec{x}^2$, $a \cdot x \equiv -a^0 \, t+\vec{a} \cdot \vec{x}$, etc) The factor multiplying $r$ on the righthand side of (\ref{scr}) is not positive definite, and this transformation connects the two patches. 
The situation is completely different for EAdS. In this case a single Poincar\'e patch has a  full conformal symmetry.

\subsection{Boundaries at $r=+\infty$ and $r=-\infty$ are connected}
\begin{figure}[h]
\begin{center}
\includegraphics[scale=0.3]{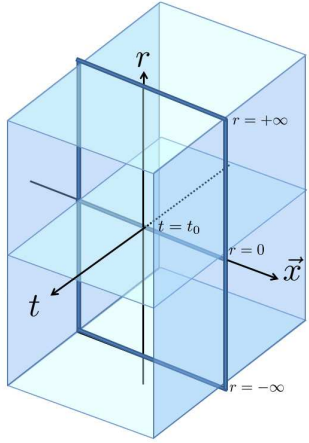} 
\end{center}
\caption{Boundaries of a pair of Poincar\'e patches except for those at $t=\pm \infty$: two boundaries at $r=\pm \infty$ are connected. Thick lines are boundaries at $t=t_0$. }  \label{PoincareBound}
\end{figure}

\hspace{5mm}
Let us study the location of the conformal boundary in the Poincar\'e coordinates. By the definition of the hyperboloid  (\ref{hyperboloideq}) it is defined by $\sum_{i=1}^d X_i^2 \rightarrow \infty$, and given by  $\rho=\pi/2$ in the global coordinates. In the Poincar\'e coordinates (\ref{Poincare}), it is given by 
\begin{equation}
\sum_{i=1}^d X_i^2 = \frac{1}{4} \, h^2 \, r^2+\ell^2 \, \vec{x}^2 \, r^2-\frac{1}{2} \, h+\frac{1}{4r^2} \rightarrow +\infty.
\end{equation}
Here $h$ is a function $h(t,\vec{x}) \equiv t^2-\vec{x}^2+\ell^2$. 
Hence the boundary of the pair of Poincar\'e patches is composed of the following hypersurfaces. 
\begin{enumerate}
\item $r \rightarrow \pm \infty$
\item $r=0$
\item $|\vec{x}| \rightarrow \infty$ with $r \neq 0$
\item $|t| \rightarrow \infty$ with $r \neq 0$
\end{enumerate} 
The union  of the above corresponds to the boundary of the global patch. 
Note that the horizon, and the spacial and even the temporal infinities are also part of the boundary. This last point is puzzling, because the conformal boundary in the global coordinates is time-like. This problem is not persued in this paper. The structure of the boundary is illustrated in Figure \ref{PoincareBound}.  Since all the parts of the boundary are connected, especially the boundaries at $r=+\infty$ and $r=-\infty$ at the same time $t$ are connected.

In the case of AdS$_2$ space the coordinates $\vec{x}$ do not exist. The boundaries $r=\pm \infty$ are connected only through the lines $t=\pm \infty$.   Hence in what follows we will consider AdS$_{d+1}$ with $d \geq 2$.

\section{Solutions to Klein-Gordon Equation in a Pair of Poincar\'e Coordinates }
\hspace{5mm}
In this section we consider a scalar field $\phi(r,t,\vec{x})$ of mass $m$ in AdS spacetime in a pair of Poincar\'e coordinates $r>0$ and $r<0$. Action integral is defined by
\begin{equation}
S_{\text{AdS}}= \int_{-\infty}^{\infty}dr \int dt \, d^{d-1}\vec{x} \, \sqrt{-g} \, (-\frac{1}{2} \, g^{\mu\nu} \, \partial_{\mu} \, 
\phi \, \partial_{\nu} \, \phi-\frac{1}{2} \, m^2 \, \phi^2). 
\end{equation}
Solution will be constructed in such a way that the fluxes across the horizon vanish and those across the boundaries at $r=\pm \infty$ cancel out. 
The resulting solution will be shown to have the following structure in a pair of Poincar\'e patches. See (\ref{modep})-(\ref{moden}). 
\begin{equation}
\phi(r,t,\vec{x}) = \left\{ \begin{array}{cc} 
\varphi_+(r,t,\vec{x})+\varphi_-(r,t,\vec{x}), \qquad (r >0), \\
S \, \varphi_+(-r,t,\vec{x})-\frac{1}{S} \, \varphi_-(-r,t,\vec{x}), \qquad (r <0).
\end{array} \right.  \label{phiform}
\end{equation}
Here $S$ is a real constant, and $\varphi_{\pm}(r,t,\vec{x})$ are functions defined for $r>0$. As $|r| \rightarrow \infty$, $\varphi_{\pm}$ behaves as $|r|^{-\Delta_{\mp}} \, \phi_{\pm}(t,\vec{x})$ ($\Delta_{\pm}>0$). Although  $\varphi_{\pm}$ generally oscillate rapidly near the horizon $r=0$, if boundary values $\phi_{\pm}$ have compact supports, we have $\varphi_{\pm} \sim |r|^{\Delta_{\pm}}$ as $r \rightarrow \pm 0$. (subsec.6.2) 
Hence, $\phi$ vanishes at $r=\pm 0$ and $r=\pm \infty$, and in a coordinate $\tilde{\rho}$ ($r=\pm e^{\tilde{\rho}}$) in stead of $r$ the solution is smooth on the entire hyperboloid.

In order to solve the equation of motion which is derived from the above action, we separate variables as 
\begin{equation}
\phi(r,t,\vec{x}) = e^{-i \omega t+i\vec{k}\cdot \vec{x}} \, \chi(r).
\end{equation}
Then $\chi(r)$ satisfies the equation
\begin{equation}
r^2 \, \partial_r^2 \, \chi+(d+1) \, r \, \partial_r \, \chi-m^2\ell^2\, \chi+(\omega^2-\vec{k}^2)\, r^{-2} \, \chi=0.   \label{chieq}
\end{equation}
Two linearly independent solutions for non-integral $\nu$ is given by 
\begin{equation}
\chi^{\pm}(r)= r^{-\frac{d}{2}} \, J_{\pm \nu}\left(\frac{\sqrt{\omega^2-\vec{k}^2}}{r}\right),   \label{chipm}
\end{equation}
where $J_{\nu}(z)$ is a Bessel function and 
\begin{equation}
\nu = \sqrt{\frac{d^2}{4}+m^2 \, \ell^2}. 
\end{equation}
We will restrict our  attention to the case where $\nu$ is real and in the range $0 < \nu <1$, because then mode functions with two different falloff behaviour can be obtained. For simplicity, we will set $\ell=1$ in what follows. 

For $\omega^2-\vec{k}^2 <0$, solutions (\ref{chipm}) blow up exponentially at either side of the horizon $r=0$ and are non-normalizable. Thus from now on we will require $\omega^2-\vec{k}^2  \geq 0$. 
In this case solutions oscillate near the horizon. The general solution to the Klein-Gordon equation can be written for $r >0$ and $r<0$ as
\begin{eqnarray}
&&\phi_{\omega,\vec{k}}(r,t,\vec{x}) \nonumber \\
&=& \left\{\begin{array}{cc}
 e^{-i \omega t+i\vec{k} \cdot \vec{x}} \, \Big( C_+(\omega,\vec{k}) \, \psi_+(r,\omega,\vec{k})+C_-(\omega,\vec{k}) \, \psi_-(r,\omega,\vec{k})\Big) & \quad (r > 0), \\
 e^{-i \omega t+i\vec{k} \cdot \vec{x}} \, \Big( \tilde{C}_+(\omega,\vec{k}) \, \psi_+(r,\omega,\vec{k})+\tilde{C}_-(\omega,\vec{k}) \, \psi_-(r,\omega,\vec{k})\Big) &  \quad (r < 0). \end{array}  \right.  \label{scalarsolution}
\end{eqnarray}
Here the mode functions are defined by 
\begin{equation}
\psi_{\pm}(r,\omega,\vec{k}) =\left\{ \begin{array}{cc} 2^{\pm \nu} \Gamma(1 \pm\nu) \, e^{\frac{i}{2}\pi \, (\frac{d}{2} \pm \nu)} \, r^{-\frac{d}{2}} \, J_{\pm \nu}\Big(\frac{\sqrt{\omega^2-\vec{k}^2}}{r}\Big) & (r>0),  \\
2^{\pm \nu} \Gamma(1 \pm\nu) \, e^{-\frac{i}{2}\pi \, (\frac{d}{2} \pm \nu)} \, (-r)^{-\frac{d}{2}} \, J_{\pm \nu}\Big(\frac{\sqrt{\omega^2-\vec{k}^2}}{-r}\Big) & (r < 0).  \end{array} \right.  \label{psi2}
\end{equation}
Because the metric (\ref{AdSmetric}) is degenerate at the horizon ($r=0)$, the equation for $\phi$ is  singular.  So, the coefficients $\tilde{C}_{\pm}$ will be connected to  $C_{\pm}$ in such a way that the fluxes are matched at the horizon and cancel out between the boundaries. 

\subsection{Klein-Gordon Norm}
\hspace{5mm}
The Klein-Gordon (KG) norm $(\phi_1,\phi_2)$ for two modes $\phi_{1,2}$ is given by\footnote{Here time $t=t_0$ is fixed. For the coordinate system in Figure \ref{fig2}, constant-$t$ hypersurfaces for $r>0$ and $r<0$ patches are not adjacent to each other at the horizon.}
\begin{eqnarray}
(\phi_1,\phi_2) &=& \int^{\infty}_{-\infty} dr \int d^{d-1}\vec{x}
\, \sqrt{-g} \, \frac{-i}{2} \, g^{tt} \, (\phi_1^{\ast} \partial_t \phi_2-\phi_2 \partial_t \phi^{\ast}_1)|_{t=t_0 \ \text{fixed}} \nonumber \\
&=&\int^{\infty}_{-\infty} dr \int d^{d-1}\vec{x} \, \frac{i}{2} \, |r|^{d-3} \, (\phi_1^{\ast} \partial_t \phi_2-\phi_2 \partial_t \phi^{\ast}_1)|_{t=t_0 \ \text{fixed}}  \label{KGnorm}
\end{eqnarray}
Although the KG current is divergenceless,  for conservation of the norm (\ref{KGnorm}), we need to impose some conditions on the solutions. 
We will show that this norm is conserved ({\em i.e.}, independent of $t_0$), if the coefficients satisfy the relations
\begin{eqnarray}
\tilde{C}_+(\omega,\vec{k}) &=& e^{i\pi \nu} \, C_+(\omega,\vec{k}) \, S \, e^{i(\alpha+\frac{\pi}{2}d)}, \label{Ctilde2C1}\\
\tilde{C}_-(\omega,\vec{k}) &=& - e^{-i\pi \nu} \, C_-(\omega,\vec{k}) \, \frac{1}{S} \, e^{i(\alpha+\frac{\pi}{2}d)}.  \label{Ctilde2C2}
\end{eqnarray}
Here $\alpha$ and $S$ are real parameters.

When solution (\ref{scalarsolution}) is substituted into the norm (\ref{KGnorm}) and $\vec{x}$ integral is performed, the norm is given by 
\begin{multline}
(\phi_1,\phi_2) =  \frac{\omega_1+\omega_2}{2} \, (2\pi)^{d-1} \,\delta^{(d-1)}(\vec{k}_1-\vec{k}_2) \, e^{i(\omega_1-\omega_2) \, t}  \\
  \cdot\Big( \int_0^{\infty} dr \, r^{d-3} \, \psi^{\ast}_1(r,\omega_1,\vec{k}_1) \, \psi_2(r,\omega_2,\vec{k}_2) \\
+    \int^0_{-\infty} dr \, (-r)^{d-3} \, \psi^{\ast}_1(r,\omega_1,\vec{k}_1) \, \psi_2(r,\omega_2,\vec{k}_2)
  \Big).  \label{norm1}
\end{multline}
Now because $\psi$ in (\ref{psi2}) solves (\ref{chieq}), $\psi_1$ and $\psi_2$ satisfy 
\begin{equation}
r^{3-d} \, \partial_r \, \big(r^{d+1} \, (\psi_1^{\ast} \, \partial_r \, \psi_2-\psi_2 \, \partial_r \, \psi_1^{\ast})\big)= (\omega_1^2-\omega_2^2-\vec{k}_1^2+\vec{k}_2^2) \, \psi_1^{\ast} \, \psi_2, 
\end{equation}
and for $\omega_1^2-\omega_2^2 \neq 0$, the norm (\ref{norm1}) is expressed in terms of boundary values.\footnote{We follow the techniques used in \cite{AM}.} 
\begin{multline}
(\phi_1,\phi_2) =  (2\pi)^{d-1} \,\delta^{(d-1)}(\vec{k}_1-\vec{k}_2) \, e^{i(\omega_1-\omega_2) \, t} \, \frac{1}{2(\omega_1-\omega_2)}  \\
 \cdot\Big(  [r^{d+1} \, (\psi_1^{\ast} \, \partial_r \, \psi_2-\psi_2 \, \partial_r \, \psi_1^{\ast})]^{\infty}_0+[(-r)^{d+1} \, (\psi_1^{\ast} \, \partial_r \, \psi_2-\psi_2 \, \partial_r \, \psi_1^{\ast})]_{-\infty}^0 \Big) 
\end{multline}
The contributions from the boundaries $r =\pm \infty$ are computed by using $J_{\nu}(z) \sim (\Gamma(\nu+1))^{-1} \, (z/2)^{\nu}$ for $z \sim 0$. The result is 
\begin{multline}
(\phi_1,\phi_2)|_{|r|=\infty} =\frac{1}{2(\omega_1-\omega_2)} \, (2\pi)^{d-1} \,\delta^{(d-1)}(\vec{k}_1-\vec{k}_2) \, e^{i(\omega_1-\omega_2) \, t}  \\
 \cdot\Big[ 2\nu \, C_+^{\ast}(\omega_1,\vec{k}_1) \, C_- (\omega_2, \vec{k}_2) \, e^{-\pi i\nu} \, \Big(\frac{\omega_1^2-\vec{k}_1^2}{ \omega_2^2-\vec{k}_2^2}   \Big)^{\frac{\nu}{2}}  
 - 2\nu \, C_-^{\ast}(\omega_1,\vec{k}_1) \, C_+ (\omega_2, \vec{k}_2) \, e^{\pi i\nu} \, \Big(\frac{\omega_2^2-\vec{k}_2^2}{ \omega_1^2-\vec{k}_1^2}   \Big)^{\frac{\nu}{2}}  \\
 + 2\nu \, \tilde{C}_+^{\ast}(\omega_1,\vec{k}_1) \, \tilde{C}_- (\omega_2, \vec{k}_2) \, e^{\pi i\nu} \, \Big(\frac{\omega_1^2-\vec{k}_1^2}{ \omega_2^2-\vec{k}_2^2}   \Big)^{\frac{\nu}{2}}  
 - 2\nu \, \tilde{C}_-^{\ast}(\omega_1,\vec{k}_1) \, \tilde{C}_+ (\omega_2, \vec{k}_2) \, e^{-\pi i\nu} \, \Big(\frac{\omega_2^2-\vec{k}_2^2}{ \omega_1^2-\vec{k}_1^2}   \Big)^{\frac{\nu}{2}} \Big].
\end{multline}
This vanishes if $C_+=\tilde{C}_+=0$ or $C_-=\tilde{C}_-=0$, {\em i.e.}, if Dirichlet or Neumann boundary condition is imposed. There is, however, another solution. This norm 
also vanishes, if the following condition is satisfied.
\begin{equation}
C^{\ast}_+(\omega_1,\vec{k}_1) \, C_-(\omega_2,\vec{k}_2)=-e^{2\pi i\nu} \, \tilde{C}^{\ast}_+(\omega_1,\vec{k}_1) \, \tilde{C}_-(\omega_2,\vec{k}_2)   \label{condCC}
\end{equation}
This new solution is possible, because a pair of Poincar\'e patches is introduced.  As will be shown in the next subsection, a flux across one boundary matches that from another.

We now turn to the contributions to the norm from the horizon. These are obtained by using the asymptotic form $J_{\nu}(z) \sim \sqrt{2/\pi z} \, \cos (z-(2\nu+1)\pi/4)$
for $z \rightarrow \infty$.  The contribution from the upper side of the horizon is given by 
\begin{multline}
(\phi_1,\phi_2)|_{r=+0} 
= \frac{1}{2(\omega_1-\omega_2)} \, e^{i(\omega_1-\omega_2)t} \, (2\pi)^{d-1} \, \delta^{(d-1)}(\vec{k}_1-\vec{k}_2)  \, \cdot\lim_{r \rightarrow 0}   \\
  \Big[ -\frac{2}{\pi} \, \big(4^{\nu} \, \Gamma(1+\nu)^2 \, N_+ + 4^{-\nu} \, \Gamma(1-\nu)^2 \, N_-\big)  \, \sin\frac{\sqrt{\omega_2^2-\vec{k}_2^2}-\sqrt{\omega_1^2-\vec{k}_1^2}}{r}  \\ 
-\frac{2}{\pi} \, \left(\frac{\omega_2^2-\vec{k}_2^2}{\omega_1^2-\vec{k}_1^2}\right)^{\frac{1}{4}} \, M_1(\omega_1,\vec{k}_1;\omega_2,\vec{k}_2) \, +\frac{2}{\pi} \, \left(\frac{\omega_1^2-\vec{k}_1^2}{\omega_2^2-\vec{k}_2^2}\right)^{\frac{1}{4}} \, M_2(\omega_1,\vec{k}_1;\omega_2,\vec{k}_2)  \\
+\frac{2}{\pi} \, \left(\frac{\omega_1^2-\vec{k}_1^2}{\omega_2^2-\vec{k}_2^2}\right)^{\frac{1}{4}} \, M_1^{\ast}(\omega_2,\vec{k}_2;\omega_1,\vec{k}_1) -\frac{2}{\pi} \, \left(\frac{\omega_2^2-\vec{k}_2^2}{\omega_1^2-\vec{k}_1^2}\right)^{\frac{1}{4}} \, M_2^{\ast}(\omega_2,\vec{k}_2;\omega_1,\vec{k}_1)  \Big],    \label{norm2}
\end{multline}
where 
\begin{equation}
N_{\pm}= C^{\ast}_{\pm}(\omega_1,\vec{k}_1) \, C_{\pm}(\omega_2,\vec{k}_2),
\end{equation}
\begin{eqnarray}
M_1(\omega_1,\vec{k}_1;\omega_2,\vec{k}_2) &=&\Gamma(1+\nu) \, \Gamma(1-\nu) \,  C_+^{\ast}(\omega_1,\vec{k}_1)C_-(\omega_2,\vec{k}_2) \, e^{-i\pi \nu} \nonumber \\
&& \cdot \cos \big(\frac{\sqrt{ \omega_1^2-\vec{k}_1^2}}{r}-\frac{2\nu+1}{4} \pi \big) \sin \big(\frac{\sqrt{ \omega_2^2-\vec{k}_2^2}}{r}-\frac{-2\nu+1}{4} \pi \big),  \\
M_2 (\omega_1,\vec{k}_1;\omega_2,\vec{k}_2) &=&\Gamma(1+\nu) \, \Gamma(1-\nu) \,  C_+^{\ast}(\omega_1,\vec{k}_1)C_-(\omega_2,\vec{k}_2) \, e^{-i\pi \nu} \nonumber \\
&& \cdot \cos \big(\frac{\sqrt{ \omega_2^2-\vec{k}_2^2}}{r}-\frac{-2\nu+1}{4} \pi \big) \sin \big(\frac{\sqrt{ \omega_1^2-\vec{k}_1^2}}{r}-\frac{2\nu+1}{4} \pi \big) 
\end{eqnarray}
To simplify $M_1$ and $M_2$, we need to use some formulae for distributions: $\sin (\Lambda x)/(\pi x) \rightarrow \delta (x)$, $\cos (\Lambda x)/(\pi x) \rightarrow 0$ for $\Lambda \rightarrow +\infty$.\cite{AM} In the limit $r \rightarrow +0$, functions $M_{1,2}$ can be simplified by using these formulae as
\begin{multline}
M_1(\omega_1,\vec{k}_1;\omega_2,\vec{k}_2)=M_2(\omega_1,\vec{k}_1;\omega_2,\vec{k}_2) \\ =\frac{1}{2} \, \Gamma(1+\nu) \, \Gamma(1-\nu) \, C_+^{\ast}(\omega_1,\vec{k}_1)C_-(\omega_2,\vec{k}_2) \,e^{-i\pi \nu}  
 \sin \frac{\sqrt{\omega_1^2-\vec{k}_1^2}-\sqrt{\omega_2^2-\vec{k}_2^2}}{r} \, \sin \frac{2\nu+1}{2} \pi.
\end{multline}
Since $(\omega_1-\omega_2)^{-1} \, \sin \frac{\sqrt{\omega_1^2-\vec{k}_1^2}-\sqrt{\omega_2^2-\vec{k}_2^2}}{r} \rightarrow  \pi \, \text{sign}(\omega_1) \,  \delta(\omega_1-\omega_2)$, those terms which contain $M_{1,2}$ all vanish, and we get 
\begin{multline}
(\phi_1,\phi_2)|_{r=+0}= \text{sign} \, (\omega_1) \, \delta(\omega_1-\omega_2) \, (2\pi)^{d-1} \, \delta^{(d-1)}(\vec{k}_1-\vec{k}_2)  \\
 \cdot \Big[ 4^{\nu} \, \Gamma(1+\nu)^2\, |C_+(\omega_1,\vec{k}_1)|^2+
4^{-\nu} \, \Gamma(1-\nu)^2\, |C_-(\omega_1,\vec{k}_1)|^2\Big]. 
\end{multline}
Contribution to the norm at the other side of the horizon, $(\phi_1,\phi_2)|_{r=-0}$, can be similarly computed.  Finally,  KG norm is  independent of $t$ and given by 
\begin{eqnarray}
(\phi_1,\phi_2) &=& \text{sign} \, (\omega_1) \, \delta(\omega_1-\omega_2) \, (2\pi)^{d-1} \, \delta^{(d-1)}(\vec{k}_1-\vec{k}_2) \, \nonumber \\
&& \cdot \Big[ 4^{\nu} \, \Gamma(1+\nu)^2(1+S^2) \, |C_+(\omega_1,\vec{k}_1)|^2+
4^{-\nu} \, \Gamma(1-\nu)^2(1+S^{-2}) \, |C_-(\omega_1,\vec{k}_1)|^2\Big]. \nonumber \\ && \label{KGconserved}
\end{eqnarray}

\subsection{Flux}
\hspace{5mm}
Since the KG norm is conserved, the fluxes must cancel or vanish at the boundaries and the horizon. Let us check this. 
One can compute the flux across the horizon from the $r >0$ patch.
\begin{equation}
J_{(+0)} =\int_{r=r_0 \rightarrow +0, \ t=t_0} \, d^{d-1}\vec{x} \, \sqrt{-g} \, g^{rr} \, \frac{-i}{2} \, (\phi_1^{\ast} \, \partial_r \, \phi_2-\phi_2 \, \partial_r \, \phi_1^{\ast})
\end{equation}
One can show that this vanishes by using $\sqrt{-g} \, g^{rr}=r^{d+1}$ and $\phi \sim r^{-(d+1)/2} \, \cos ( \cdots )$. A calculation similar to that used in deriving (\ref{KGconserved}) must be done. Similarly, the flux at $r=-0$ also vanishes. The fluxes at $r=+\infty$, however,  does not vanish. 
It is given by 
\begin{multline}
J_{(+\infty)}=-i\nu \, (2\pi)^{d-1} \, \delta^{(d-1)}(\vec{k}_1-\vec{k}_2) \, e^{i(\omega_1-\omega_2)t} \cdot \\
\Big( \left(\frac{\omega_1^2-\vec{k}_1^2}{\omega_2^2-\vec{k}_2^2}\right)^{\frac{\nu}{2}}e^{-i\pi \nu} \, C_+^{\ast}(\omega_1,\vec{k}_1) \, C_-(\omega_2,\vec{k}_2) 
- \left(\frac{\omega_2^2-\vec{k}_2^2}{\omega_1^2-\vec{k}_1^2}\right)^{\frac{\nu}{2}}e^{i\pi \nu} \, C_-^{\ast}(\omega_1,\vec{k}_1) \, C_+(\omega_2,\vec{k}_2) \Big). 
\end{multline}
This takes forms of interference terms between the two kinds of modes. 
By using (\ref{condCC}) one can show that this is canceled by the out-going flux at $r=-\infty$. 
\begin{multline}
J_{(-\infty)}=+i\nu \, (2\pi)^{d-1} \, \delta^{(d-1)}(\vec{k}_1-\vec{k}_2) \, e^{i(\omega_1-\omega_2)t} \cdot \\
\Big( -\left(\frac{\omega_1^2-\vec{k}_1^2}{\omega_2^2-\vec{k}_2^2}\right)^{\frac{\nu}{2}}e^{i\pi \nu} \, \tilde{C}_+^{\ast}(\omega_1,\vec{k}_1) \, \tilde{C}_-(\omega_2,\vec{k}_2) 
+ \left(\frac{\omega_2^2-\vec{k}_2^2}{\omega_1^2-\vec{k}_1^2}\right)^{\frac{\nu}{2}}e^{-i\pi \nu} \, \tilde{C}_-^{\ast}(\omega_1,\vec{k}_1) \, \tilde{C}
_+(\omega_2,\vec{k}_2) \Big)
\end{multline}

The above results might seem useless, because the two boundaries in Figure \ref{fig2} appear to be infinitely separated. As mentioned in subsec 2.4, however, in AdS$_{d+1}$ space with $d \geq 2$,  
the two boundaries at $r=+\infty$ and $r=-\infty$ are connected.  In this way the total flux computed on the boundaries $r=\pm \infty$ cancels out at any time $t$.

To summarize,  normalizable modes in the pair of Poincar\'e patches are given by 
\begin{eqnarray}
\phi_{\omega,\vec{k}}(r,t,\vec{x}) &=& e^{-i\omega t+i\vec{k} \cdot \vec{x}} \nonumber \\
&& \cdot [ C_+(\omega,\vec{k}) \, 2^{\nu} \, \Gamma(1+\nu) \, e^{\frac{i}{2}\pi (\frac{d}{2}+\nu)} \, r^{-\frac{d}{2}} \, J_{\nu}(\sqrt{\omega^2-\vec{k}^2}/r) \nonumber \\
&& +C_-(\omega,\vec{k}) \, 2^{-\nu} \, \Gamma(1-\nu) \, e^{\frac{i}{2}\pi (\frac{d}{2}-\nu)} \, r^{-\frac{d}{2}} \, J_{-\nu}(\sqrt{\omega^2-\vec{k}^2}/r)] \label{modep}
\end{eqnarray}
for $r >0$, and 
\begin{eqnarray}
\phi_{\omega,\vec{k}}(r,t,\vec{x}) &=& e^{-i\omega t+i\vec{k} \cdot \vec{x}} \, e^{i\alpha}  \nonumber \\
&& \cdot [ S \, C_+(\omega,\vec{k}) \, 2^{\nu} \, \Gamma(1+\nu) \, e^{\frac{i}{2}\pi (\frac{d}{2}+\nu)} \, (-r)^{-\frac{d}{2}} \, J_{\nu}(-\sqrt{\omega^2-\vec{k}^2}/r) \nonumber \\
&& -\frac{1}{S} \, C_-(\omega,\vec{k}) \, 2^{-\nu} \, \Gamma(1-\nu) \, e^{\frac{i}{2}\pi (\frac{d}{2}-\nu)} \, (-r)^{-\frac{d}{2}} \, J_{-\nu}(-\sqrt{\omega^2-\vec{k}^2}/r)] \nonumber \\ && \label{moden}
\end{eqnarray}
for $r <0$. Note that these mode functions are rapidly oscillating and blowing up near the horizon $r=0$ like $\sim r^{\frac{1-d}{2}} \, \cos (\sqrt{\omega^2-\vec{k}^2}/r-(\pm 2 \nu+1)\pi/4)$. However, this very rapid oscillation actually makes the mode functions cancel out and vanish at the horizon. We will show in sec.6 that the solution (\ref{phiplus}) to the boundary-value problem, constructed by smearing these mode functions by source functions which have compact supports, has a milder behaviour $\phi \sim |r|^{\Delta_{\pm}}$ for $r \rightarrow 0$, where $\Delta_{\pm}=d/2 \pm\nu$. By means of the coordinate \footnote{This variable $\tilde{\rho}$ is different from $\rho$ of the global coordinates (\ref{globalcoord}).} $\tilde{\rho} \equiv  \log |r|$ this can be written as $\phi \sim e^{\Delta_{\pm} \tilde{\rho}}$, and $\phi$ asymptotes to zero exponentially near the horizon $\tilde{\rho}=-\infty$. In this sense, the mode functions are smoothly connected at the horizon. 

\subsection{Conservation of Energy}
\hspace{5mm}
In AdS$_{d+1}$ there is a time-like Killing vector and by contracting this with a stress-energy tensor, a formally conserved energy can be defined. To obtain an exactly conserved energy, one needs to show that the energy-flux vanishes or cancels at the horizon and boundaries. In AdS space the Riemann scalar $R$ is constant, $d(d+1)$ (in units $\ell=1$), and a coupling $R \, \phi^2$ is equivalent to a mass term. Hence we may replace the mass squared $m^2$ by $m^2 -\xi \, d(d+1)+\xi \, R $ in the action.  Here $\xi$ is a constant and the conformal coupling corresponds to $\xi=\xi_c \equiv -(d-1)/(4d)$. We will leave $\xi$ as a free parameter and fix its value below.\footnote{In \cite{BF} it was shown that in the global coordinates of AdS space, energy of either Dirichlet or Neumann mode is conserved by choosing stress-tensor with a conformal coupling $\xi=\xi_c$.  } 

The stress-energy tensor is, after substitution of the solution into the equation of motion, given by
\begin{eqnarray}
T_{\mu\nu} &=& (1+2\xi) \, \partial_{\mu} \phi \, \partial_{\nu} \phi+2\xi \, \phi \, \nabla_{\mu}\nabla_{\nu}\phi+\big(2\xi-\frac{1}{2}\big) \, m^2 \, g_{\mu\nu} \, \phi^2 \nonumber \\
&&-\frac{1}{2}\, (1+4\xi) \, g_{\mu\nu}\, \phi \, \nabla^{\lambda} \, \nabla_{\lambda}\, \phi-d^2 \ \xi \, g_{\mu\nu} \, \phi^2.
\end{eqnarray}
Energy flux 
\begin{equation}
\int_{t, r \, \text{fixed}} d^{d-1}\vec{x} \sqrt{-g} \, g^{rr} \, T_{rt}
\end{equation}
can be calculated as in the previous subsection for the particle number flux. $T_{rt}$ is given by
\begin{equation}
T_{rt}= (1+2\xi) \, \partial_r \, \phi \, \partial_t \, \phi+2 \, \xi \, \phi \, \partial_r \, \partial_t \, \phi-\frac{2\xi}{r} \, \phi \, \partial_t \, \phi
\end{equation} 
and it is easily shown that the fluxes at $r=+0$ and $r=-0$ vanish. 
It turns out, however,  that for general $\xi$, the energy-fluxes at $r= \pm \infty$ contain an infinity $|r|^{2\nu}$ associated with the modes $\psi_-$. This infinity can be removed by fine tuning $\xi$.
\begin{equation}
\xi=\frac{2\nu-d}{4(d-2\nu+1)}
\end{equation}
Interestingly, at $\nu=\frac{1}{2}$, this agrees with the conformal value $\xi_c$ presented above. There still remain finite (${\cal O}(r^0)$) fluxes at the two boundaries. It can, however, be shown that the remaining fluxes at $r=+\infty$ and $r=-\infty$ cancel out completely by using (\ref{condCC}), exactly as in the particle number flux. Hence the energy associated with both kinds of modes $\psi_{\pm}$ is conserved in the pair of Poincar\'e patches.

\section{Mode Expansion of $\phi$ and Canonical Commutation Relations}
\hspace{5mm}
In this section we will perform canonical quantization of a free scalar field in AdS$_{d+1}$ in a pair of Poincar\'e coordinates. We use mode expansions (\ref{modep}) and (\ref{moden}). By replacing the coefficients $C_{+,-}$ by annihilation and creation operators, and integrating over $\omega$ and $\vec{k}$, we obtain the following operator. 
\begin{eqnarray}
\Phi(r,t,\vec{x}) &=& \int^{\infty}_{-\infty} d^{d-1}\vec{k} \int^{\infty}_{|\vec{k}|}d\omega \nonumber \\
&&\cdot \Big[ e^{-i\omega t+i\vec{k} \cdot \vec{x}} \Big(a_+(\omega,\vec{k}) \, \hat{\psi}_+(r,\omega,\vec{k})+
a_-(\omega,\vec{k}) \, \hat{\psi}_-(r,\omega,\vec{k}) \Big) \nonumber \\
&& + e^{i\omega t-i\vec{k} \cdot \vec{x}} \Big(a_+^{\dagger}(\omega,\vec{k}) \, \hat{\psi}_+^{\ast}(r,\omega,\vec{k})+
a_-^{\dagger}(\omega,\vec{k}) \, \hat{\psi}_-^{\ast}(r,\omega,\vec{k}) \Big) \Big] \label{modephi}
\end{eqnarray}
The integration region is restricted to $|\vec{k}|\leq \omega$. This operator is defined for both $r>0$ and $r<0$. The functions $\hat{\psi}_{\pm}$ are obtained by slightly modifying $\psi_{\pm}$, and given by
\begin{equation}
\hat{\psi}_+(r,\omega,\vec{k}) = \left\{ \begin{array}{cc}
      2^{\nu} \, \Gamma(1+\nu) \, r^{-\frac{d}{2}} \, J_{\nu}\big(\sqrt{\omega^2-\vec{k}^2}/r  \big), & (r>0) \\ S \, e^{i\alpha} \cdot 
2^{\nu} \, \Gamma(1+\nu) \, (-r)^{-\frac{d}{2}} \, J_{\nu}\big(-\sqrt{\omega^2-\vec{k}^2}/r  \big), & (r<0) \end{array} \right. \label{psip}
\end{equation}
\begin{equation}
\hat{\psi}_-(r,\omega,\vec{k}) = \left\{ \begin{array}{cc}
      2^{-\nu} \, \Gamma(1-\nu) \, r^{-\frac{d}{2}} \, J_{-\nu}\big(\sqrt{\omega^2-\vec{k}^2}/r  \big), & (r>0) \\ - \frac{1}{S} \, e^{i\alpha} \cdot 
2^{-\nu} \, \Gamma(1-\nu) \, (-r)^{-\frac{d}{2}} \, J_{-\nu}\big(-\sqrt{\omega^2-\vec{k}^2}/r  \big). & (r<0) \end{array} \right. \label{psim}
\end{equation}

This operator and its canonical conjugate momentum $\Pi(r,t,\vec{x})=|r|^{d-3} \, \partial_t \, \Phi$ must satisfy the canonical commutation relations: $[\Phi(r,t,\vec{x}), \Pi(r',t,\vec{x}')]=i \, \delta(r-r') \, \delta^{(d-1)}(\vec{x}-\vec{x}')$, $[\Phi(r,t,\vec{x}), \Phi(r',t,\vec{x}')]=0$ and $[\Pi(r,t,\vec{x}), \Pi(r',t,\vec{x}')]=0$.  It can be shown that this is achieved by setting $\alpha=0$ or $\pi$ and imposing the following commutators. (Other commutators are vanishing.)
\begin{eqnarray}
\ [ a_+(\omega,\vec{k}), a^{\dagger}_+(\omega', \vec{k}')] &=& 
\frac{1}{2^{1+2\nu} \, (2\pi)^{d-1} \, \Gamma(1+\nu)^2} \, \frac{1}{1+S^2} \, \delta(\omega-\omega') \, \delta^{(d-1)}(\vec{k}-\vec{k}'), \nonumber \\
\ [ a_-(\omega,\vec{k}), a^{\dagger}_-(\omega', \vec{k}')] &=& 
\frac{1}{2^{1-2\nu} \, (2\pi)^{d-1} \, \Gamma(1-\nu)^2} \, \frac{S^2}{1+S^2} \, \delta(\omega-\omega') \, \delta^{(d-1)}(\vec{k}-\vec{k}') \nonumber \\
&& \label{aad}
\end{eqnarray}
Because $e^{i\alpha}$ is multiplied by $S$ or $1/S$  in (\ref{psip}), (\ref{psim}), we can set $\alpha=0$ by allowing $S$ to take positive or negative values. 

The role of parameter $S$ is to specify the relative magnitude of the mode functions (\ref{psip})-(\ref{psim}) in the two patches. One can replace $\psi_-$  by $S^{-1} \, \tilde{\psi}_-$ and $a_-$ by $S \, \tilde{a}_-$ without changing the form of (\ref{modephi}). Then, $S$-dependences of $[a_+, a_+^{\dagger}]$ and $[\tilde{a}_-,\tilde{a}_-^{\dagger}]$ become the same: $1/(1+S^2)$. If one sets $S=0$, then the mode $\tilde{\psi}_-$ is  quantized only in the patch with $r<0$, while  $\psi_+$ is  quantized only in the $r>0$ patch. 
At present we do not have an argument to determine $S$, and in this paper we will leave the value of $S$ undetermined.

First let us consider $[\Phi(r,t,\vec{x}), \Phi(r',t,\vec{x}')]$. This is given as
\begin{eqnarray}
&& [\Phi(r,t,\vec{x}), \Phi(r',t,\vec{x}')] \nonumber \\
&=& \int d^{d-1}\vec{k} \int^{\infty}_{|\vec{k}|} d\omega (2\pi)^{1-d} \, (1+S^2)^{-1}  \,   e^{i \vec{k} \cdot (\vec{x}-\vec{x}')} \nonumber \\
&& \cdot \Big[2^{-1-2\nu} \frac{1}{\Gamma(1+\nu)^2} \, \{ \hat{\psi}_+(r,\omega,\vec{k}) \, \hat{\psi}_+^{\ast}(r',\omega,\vec{k})-\hat{\psi}_+^{\ast}(r,\omega,-\vec{k}) \, \hat{\psi}_+(r',\omega,-\vec{k})\} \nonumber \\
&& +2^{-1+2\nu} \frac{S^2}{\Gamma(1-\nu)^2} \, \{ \hat{\psi}_-(r,\omega,\vec{k}) \, \hat{\psi}_-^{\ast}(r',\omega,\vec{k})-\hat{\psi}_-^{\ast}(r,\omega,-\vec{k}) \, \hat{\psi}_-(r',\omega,-\vec{k})\} \Big]. \nonumber \\&&
\end{eqnarray}
For $rr' >0$, terms on the right hand side cancel out completely. For $r>0$ and $r' <0$, we have
\begin{multline}
  [\Phi(r,t,\vec{x}), \Phi(r',t,\vec{x}')]  \\= \int d^{d-1}\vec{k} \int^{\infty}_0d\mu \frac{\mu}{\sqrt{\vec{k}^2+\mu^2}} \, e^{i\vec{k} \cdot (\vec{x}-\vec{x}')} \frac{S}{2(1+S^2)} \, (2\pi)^{1-d}
 \\
 \cdot (e^{-i\alpha}-e^{i\alpha}) \, \Big[ J_{\nu}(\frac{\mu}{r}) \, J_{\nu}(\frac{\mu}{-r'})+J_{-\nu}(\frac{\mu}{r}) \, J_{-\nu}(\frac{\mu}{-r'})    \Big] \, (-rr')^{-\frac{d}{2}}.
\end{multline}
Here we set $\omega=\sqrt{\vec{k}^2+\mu^2}$ and integration over $\omega$ is replaced by that over $\mu$. This vanishes, if $e^{i\alpha}=\pm 1$. Similar result is obtained for $r<0$ and $r' >0$.
By a similar analysis it can be shown that $[\Pi(r,t,\vec{x}), \Pi(r',t,\vec{x}')]=0$, if  $e^{i\alpha}=\pm 1$.

Next we turn to $[\Phi(r,t,\vec{x}), \Pi(r',t,\vec{x}')]$. In this case we have
\begin{multline}
 [\Phi(r,t,\vec{x}), \Pi(r',t,\vec{x}')] 
= \int d^{d-1}\vec{k} \int^{\infty}_{|\vec{k}|} d\omega (2\pi)^{1-d} \, (1+S^2)^{-1}  \,   e^{i \vec{k} \cdot (\vec{x}-\vec{x}')} \, i\omega \, |r'|^{d-3}  \\
 \cdot \Big[2^{-1-2\nu} \frac{1}{\Gamma(1+\nu)^2} \, \{ \hat{\psi}_+(r,\omega,\vec{k}) \, \hat{\psi}_+^{\ast}(r',\omega,\vec{k})+\hat{\psi}_+^{\ast}(r,\omega,-\vec{k}) \, \hat{\psi}_+(r',\omega,-\vec{k})\}  \\
 +2^{-1+2\nu} \frac{S^2}{\Gamma(1-\nu)^2} \, \{ \hat{\psi}_-(r,\omega,\vec{k}) \, \hat{\psi}_-^{\ast}(r',\omega,\vec{k})+\hat{\psi}_-^{\ast}(r,\omega,-\vec{k}) \, \hat{\psi}_-(r',\omega,-\vec{k})\} \Big].  
\end{multline}
For $rr'>0$, this yields
\begin{multline}
 [\Phi(r,t,\vec{x}), \Pi(r',t,\vec{x}')]  \\
= \int d^{d-1}\vec{k} \int^{\infty}_0 d\mu (2\pi)^{1-d} \, (1+S^2)^{-1}  \,   e^{i \vec{k} \cdot (\vec{x}-\vec{x}')} \, i\mu \, |r'|^{d-3}   \\
 \cdot  (rr')^{-\frac{d}{2}} \, \Big[ J_{\nu}\big(\frac{\mu}{r} \big)\, J_{\nu}\big(\frac{\mu}{r'}\big) +S^2 \, J_{-\nu}\big(\frac{\mu}{r}\big) \, J_{-\nu}\big(\frac{\mu}{r'}\big)\Big]. \label{phipi}
\end{multline}
By using a formula
\begin{equation}
 \int^{\infty}_0 dx \, x \, J_{\nu}(ax) \, J_{\nu}(bx)  
=\frac{1}{a^2-b^2} \ \Big[x(J_{\nu}(ax)J'_{\nu}(bx)-J_{\nu}(bx)J'_{\nu}(ax))\Big]_0^{\infty}
\label{Lommel1}
\end{equation}
and the identities for distributions used in the previous section, it can be shown that (\ref{phipi}) agrees with
$i \delta(r-r') \delta^{(d-1)}(\vec{x}-\vec{x}')$. For $r>0$ and $r' <0$ we have
\begin{multline}
 [\Phi(r,t,\vec{x}), \Pi(r',t,\vec{x}')] \\
= \int d^{d-1}\vec{k} \int^{\infty}_0 d\mu (2\pi)^{1-d} \, \frac{S}{2(1+S^2)} \,   e^{i \vec{k} \cdot (\vec{x}-\vec{x}')} \, i\mu 
\, |r'|^{d-3}   \\
 \cdot (e^{-i\alpha}+e^{i\alpha}) (-rr')^{-\frac{d}{2}} \, \Big[ J_{\nu}\big(\frac{\mu}{r} \big)\, J_{\nu}\big(\frac{\mu}{-r'}\big) - J_{-\nu}\big(\frac{\mu}{r}\big) \, J_{-\nu}\big(\frac{\mu}{-r'}\big)\Big].
\end{multline}
This vanishes due to (\ref{Lommel1}). The commutators with $r<0$ and $r' >0$ also vanish. 

\section{Wightman Function}
\hspace{5mm}
In this section we will compute Wightman function for a scalar field in AdS$_{d+1}$ space. 
\begin{equation}
G(r,t,\vec{x}; r', t', \vec{x}') = \langle 0| \Phi(r,t,\vec{x}) \,\Phi(r',t',\vec{x}')|0 \rangle
\end{equation} 
Here $|0 \rangle$ is a vacuum which is annihilated by $a_+$ and $a_-$. 

By using the mode expansion (\ref{modephi}) and the commutation relations (\ref{aad}), Wightman function is given by 
\begin{multline}
G(r,t,\vec{x}; r', t', \vec{x}') =  G_+(r,t,\vec{x}; r', t', \vec{x}')+G_-(r,t,\vec{x}; r', t', \vec{x}'),  \\
G_{\pm} = \int \frac{d^{d-1}\vec{k}}{(2\pi)^{d-1}} \int^{\infty}_{|\vec{k}|} d\omega \frac{1}{2(1+S^2)} \, 
e^{-i\omega (t -t') +i \vec{k} \cdot (\vec{x}-\vec{x}')}  \\
 \cdot  \frac{1}{2^{\pm 2\nu} \, \Gamma(1 \pm \nu)^2} \, \hat{\psi}_{\pm}(r,\omega,\vec{k}) \, \hat{\psi}_{\pm}^{\ast}(r',\omega,\vec{k}),  \label{Dpm}
\end{multline}
$G_{\pm}$ can be expressed as integrals (\ref{Wight}) of a flat-space Wightman function integrated over a mass parameter $\mu$.
We will display the results for space-like separation of the plane ($\mathbb{E}^{d-1,1}$) coordinates.  
\begin{equation}
(x-x')^2\equiv-(t-t')^2+(\vec{x}-\vec{x}')^2 >0,
\end{equation} 
From the structure of the mode functions (\ref{psip}), (\ref{psim}),  we have 
\begin{eqnarray}
G_+(r,t,\vec{x}; r', t', \vec{x}')  &=& S^{\theta(-r)+\theta(-r')} \, G_+(|r|,t,\vec{x};| r'|, t', \vec{x}'), \label{GG1} \\
G_-(r,t,\vec{x}; r', t', \vec{x}')  &=& (-S)^{-\theta(-r)-\theta(-r')} \, G_-(|r|,t,\vec{x};| r'|, t', \vec{x}'). \label{GG2}
\end{eqnarray}

By using some mathematical formulae in, for example \cite{Grad}, we can show that
\begin{eqnarray}
&&G_+(r,t,\vec{x}; r', t', \vec{x}') \nonumber \\
&= & \frac{S^{\theta(-r)+\theta(-r')}}{1+S^2}   \, \frac{\Gamma(\frac{d}{2}+\nu)}{2\pi^{\frac{d}{2}} \, \Gamma(1+\nu)} \, P^{\frac{d}{2}+\nu} \, _2F_1\big(\frac{d}{2}+\nu, \frac{1}{2}+\nu,1+2\nu;-4P\big), \label{Dpulus}
\end{eqnarray}
\begin{multline}
G_-(r,t,\vec{x}; r', t', \vec{x}')  \\
=\frac{(-S)^{\theta(r)+\theta(r')}}{1+S^2} \frac{\Gamma(\frac{d}{2}-\nu)}{2\pi^{\frac{d}{2}} \, \Gamma(1-\nu)} \, P^{\frac{d}{2}-\nu} \, _2F_1\big(\frac{d}{2}-\nu, \frac{1}{2}-\nu,1-2\nu;-4P\big). 
\label{Dminus}
\end{multline}
Here $_2F_1(a,b,c;z)$ is a hypergeometric function, and $\theta(x)$ is a step function ($\theta(x)=1$ for $x>0$ and 0 for $x<0$).  $P$ is defined by
\begin{equation}
P= \frac{1}{|rr'|} \, \frac{1}{\Big(\frac{1}{|r|}-\frac{1}{|r'|}\Big)^2+(x-x')^2} \label{P}
\end{equation}
and related to the chordal distance $\sigma \equiv X \cdot X'+1$ by 
\begin{equation}
P^{-1}= \left\{ \begin{array}{cc} 
-2\sigma &  (  rr'>0),  \\
 2(\sigma-2) & (  rr' <0).  \end{array} \right.
\end{equation}
Note that $\sigma=0$ for $X=X'$, and $\sigma=2$ for $X=-X'$. 
Hence $P^{-1}$ vanishes either if the points coincide $X=X'$ $(rr' >0)$, or if they are antipodal to each other, $X=-X'$ $(rr' <0)$.
%\footnote{Since the metric is indefinite,  $X \cdot X'=\mp 1$ and $X^2=X'^2=-1$ do not necessarily mean $X=\pm X'$. $P^{-1}$ vanishes on the light-cone of these points. } 
The hypergeometric functions in (\ref{Dpulus}) and (\ref{Dminus}) can be singular at $-4P=0,1,\infty$. As discussed above, condition $P=\infty$, which is equivalent to 
\begin{equation}
\Big(\frac{1}{|r|}-\frac{1}{|r'|}\Big)^2+(x-x')^2=0,   \label{charge1}
\end{equation}
is satisfied for $X= \pm X'$ (coincident and antipodal points).\footnote{It can be shown  by using (\ref{Dpulus}-\ref{Dminus}) that singularities at $X=-X'$ cancel out between $G_+$ and $G_-$.}  
A singularity at $P=-1/4$ occurs for 
\begin{equation}
\Big(\frac{1}{|r|}+\frac{1}{|r'|}\Big)^2+(x-x')^2=0.   \label{charge2}
\end{equation}
These singularities (\ref{charge1}) and (\ref{charge2}) are associated with a real charge and its image.\cite{Gibbons}  If $d \geq 2$, $P=0$ is not a singularity due to the pre-factors $P^{\frac{d}{2} \pm \nu}$ in (\ref{Dpulus}) and (\ref{Dminus}).  
%Although the Wightman function has singularity with an antipode, there is nothing acausal, since the scalar field $\Phi$ satisfies the canonical commutation relations and its commutator vanishes for space-like separations. 
The result (\ref{Dpulus}) is derived in Appendix A. When $r'$ is sent to infinity, the above functions approach the bulk-boundary propagators: $G_{\pm} \sim \text{const} \cdot \, |r'|^{-\Delta_{\pm}} \, ((1+r^2 (x-x')^2)/|r|)^{-\Delta_{\pm}}$. 
For null and time-like separation ($P^{-1} \leq 0$), $G$ is given by analytically continuing the above result by $i\epsilon$ prescription $t-t' \rightarrow t-t'-i\epsilon$. 
Feynman propagator $iG_F$ is obtained from $G$ by replacement $P^{-1} \rightarrow P^{-1}+i\epsilon$.  
Feynman propagator of a scalar field in a Poincar\'e patch of AdS space with a single type of modes  was obtained in \cite{BFDG}.

\section{AdS/CFT Correspondence}
\hspace{5mm}
In the preceding sections we have learnt that a general solution $\phi$ to the K-G equation 
in a pair of Poincar\'e patches has the structure (\ref{phiform}).
%\begin{equation}
%\phi(r,t,\vec{x}) = \left\{ \begin{array}{cc} 
%\varphi_+(r,t,\vec{x})+\varphi_-(r,t,\vec{x}), \qquad (r >0), \\
%S \, \varphi_+(-r,t,\vec{x})-\frac{1}{S} \, \varphi_-(-r,t,\vec{x}), \qquad (r <0).
%\end{array} \right.  \label{phiform}
%\end{equation}
%Here $\varphi_+(r,t,\vec{x})$ and $\varphi_-(r,t,\vec{x})$ are defined only for $r>0$. 
As discussed at the end of sec.3, $\varphi_{\pm}$ in a coordinate $\tilde{\rho} \equiv \log |r|$ effectively asymptote to zero exponentially near the horizon $\tilde{\rho} \rightarrow -\infty$, and $\phi$ is smooth at the horizon, even if $\varphi_{\pm}$ are multiplied by $S$ and $-1/S$ for $r<0$. 
As we will see, this structure imposes some constraints on the boundary conditions for $\phi$ at $r=\pm \infty$. The above relations remind us of the connection between Fourier series expansion in an interval $(-\pi, \pi)$, and sinusoidal and cosinusoidal Fourier series expansions in a half interval $(0,\pi)$.\cite{AIS}\cite{BF}\cite{BKL} In that case the sinusoidal one is odd under the reflection and the cosinusoidal one is even. Here this correspondence is modified by the extra factors $S$ and $1/S$.

\subsection{  Wick rotation}
\hspace{5mm}
In what follows we will switch to Euclidean Anti-de Sitter space (EAdS$_{d+1}$) by Wick rotation. In contrast to AdS$_{d+1}$, the quadric in $\mathbb{E}^{d+1,1}$ is composed of two hyperbolic spaces $H^{d+1}$ (disconnected balls $B_{d+1}$). Each piece has $r>0$ and $r<0$, respectively. One of the two {\em is} EAdS$_{d+1}$. Hence one usually quantizes a scalar field in a single Poincar\'e patch. When the entire Lorentzian AdS space is considered, however, one cannot go from a Lorentzian signature to a Euclidean one, and then come back through analytical continuation. Our primary concern is to study a scalar field theory in Lorentzian AdS space, not in EAdS. We perform Wick rotations in order to make integrals which contain products of bulk-boundary propagators well-defined, when the UV divergence is regularized by cutoff $|r|=$ finite. Hence, in what follows, we will consider both pieces of the hyperbolic spaces, and glue together the two half spaces at the horizon, which is also part of the boundary.  
Then, we assume that the structure of the solution (\ref{phiform}) is the same after Wick rotation, although the topology of the spacetime has changed by Wick rotation. 
The coordinates on the boundary will be denoted as $\vec{y}$ instead of $(\tau=it,\vec{x})$. 
The bulk action integral for the scalar field is given by
\begin{equation}
I_0 = \int_{-\infty}^{\infty} dr \, \int d^d y \sqrt{g} \, \Big(-\frac{1}{2} \, \frac{1}{\sqrt{g}}\, \phi \, \partial_{\mu} (\sqrt{g} \, g^{\mu\nu} \, \partial_{\nu}\phi) +\frac{1}{2} \, m^2 \, \phi^2\Big). \label{bulkaction}
\end{equation}
Here note that this action has an asymmetric form. This is different from the action in ordinary form by surface terms. This is arranged so that the on-shell value of $I_0$ vanishes.\cite{ads4} Surface action integrals $I_{\pm}$ will also be introduced later, and the total action is $I=I_0+I_++I_-$. The metric tensor is given by 
\begin{equation}
ds_{\small E}^2= g_{\mu\nu} \, dx^{\mu}dx^{\nu}= r^{-2} dr^2+r^2 \, (d\vec{y})^2.
\end{equation}

The equation of motion for $\phi$ in the bulk has solutions of a form $\phi \sim r^{-(d-\Delta)}$ as $|r| \rightarrow \infty$. There are two values of $\Delta$
\begin{equation}
\Delta=\Delta_{\pm} \equiv \frac{d}{2} \pm \nu, \qquad 
\nu = \sqrt{\left(\frac{d}{2}\right)^2+m^2}.
\end{equation}
BF bound\cite{BF} is given by $m^2 \geq -\frac{d^2}{4}$. When $\Delta_-$ satisfies the unitarity bound $\Delta_- \geq \frac{d-2}{2}$, $\nu$ will be in the range $0 < \nu <1$.\footnote{Values $\nu=0, 1$ are not considered in this paper.}  If $\nu$ is in this range, there are two scalar operators ${\it O}_+$, ${\it O}_-$ with scaling dimensions $\Delta_+$, $\Delta_-$ in the boundary CFT. Discussion in this paper will be restricted to this case. Then $ \Delta_- < \frac{d}{2} < \Delta_+$.

\subsection{Green Functions and Solutions to Boundary-Value Problem}
\hspace{5mm}
Euclidean Green function $G_E(r,y;r',y')$ is obtained from Feynman propagator $iG_F$ ($y=(\tau, \vec{x})$) by the relation 
\begin{equation}
G_E(r,\tau, \vec{x};r',\tau',\vec{x}')=i G_F(r,-i\tau, \vec{x};r',-i\tau',\vec{x}')
\end{equation}
The bulk-boundary Green functions are given by\cite{ads3}\cite{FMMR}
\begin{equation}
K_{\Delta_{\pm}}(\vec{y},\vec{y} \, ',r) = \frac{\Gamma(\Delta_{\pm})}{\pi^{d/2} \, \Gamma(\Delta_{\pm}-\frac{d}{2})} \, \Big( \frac{r}{1+r^2 \, (\vec{y}-\vec{y} \, ')^2}\Big)^{\Delta_{\pm}}.
\label{bulkboundary}
\end{equation}
Near the boundary ($r \rightarrow \infty$), these have the asymptotics.
\begin{equation}
K_{\Delta_{\pm}}(\vec{y},\vec{y} \, ',r) \rightarrow r^{-(d-\Delta_{\pm}) } \, \delta^{(d)}(\vec{y}-\vec{y} \, ')+r^{-\Delta_{\pm}} \, \frac{\Gamma(\Delta_{\pm})}{\pi^{d/2} \, \Gamma(\Delta_{\pm}-\frac{d}{2})} \, \frac{1}{|\vec{y}-\vec{y} \, '|^{2\Delta_{\pm}}}+\dots  \label{Kasymp}
\end{equation}
Due to (\ref{Dpulus})-(\ref{Dminus}) these are related to $G_{\pm}$ by 
\begin{equation}
G_{\pm}(|r|,-i\tau,\vec{x};|r'|,-i\tau',\vec{x}')= \frac{1}{\pm 2(1+S^{\pm 2})\nu} \, \frac{1}{|r'|^{\Delta_{\pm}}} \, K_{\Delta_{\pm}}(\vec{y},\vec{y} \, ',|r|),  \quad r' \rightarrow \infty. \label{GK}
\end{equation}
Then  we can write down the general solution to the Klein-Gordon equation in the pair of Poincar\'e patches for EAdS$_{d+1}$.  
\begin{eqnarray}
\phi(r,\vec{y}) &=& \frac{1}{1+S^2} \, \Big[\int d^d \vec{y} \, ' \, K_{\Delta_+}(\vec{y},\vec{y} \, ',r) \, \phi_+(\vec{y} \, ')+
S \, \int d^d \vec{y} \, ' \, K_{\Delta_+}(\vec{y},\vec{y} \, ',r) \, \bar{\phi}_+(\vec{y} \, ') \Big]  \nonumber \\
&& +\frac{1}{1+S^2} \, \Big[S^2 \, \int d^d \vec{y} \, ' \, K_{\Delta_-}(\vec{y},\vec{y} \, ',r) \, \phi_-(\vec{y} \, ')-
S \, \int d^d \vec{y} \, ' \, K_{\Delta_-}(\vec{y},\vec{y} \, ',r) \, \bar{\phi}_-(\vec{y} \, ') \Big]   \nonumber \\ && (r > 0)  \label{general1}
\end{eqnarray}
and  
\begin{eqnarray}
\phi(r,\vec{y}) &=& \frac{1}{1+S^2} \, \Big[S \, \int d^d \vec{y} \, ' \, K_{\Delta_+}(\vec{y},\vec{y} \, ',-r) \, \phi_+(\vec{y} \, ')+
S^2 \, \int d^d \vec{y} \, ' \, K_{\Delta_+}(\vec{y},\vec{y} \, ',-r) \, \bar{\phi}_+(\vec{y} \, ') \Big]  \nonumber \\
&& +\frac{1}{1+S^2} \, \Big[-S \, \int d^d \vec{y} \, ' \, K_{\Delta_-}(\vec{y},\vec{y} \, ',-r) \, \phi_-(\vec{y} \, ')+ \int d^d \vec{y} \, ' \, K_{\Delta_-}(\vec{y},\vec{y} \, ',-r) \, \bar{\phi}_-(\vec{y} \, ') \Big]   \nonumber \\ && (r <0).  \label{general2}
\end{eqnarray}
Here $\phi_{\pm}(\vec{y})$ and $\bar{\phi}_{\pm}(\vec{y})$ are boundary conditions at $r=+\infty$ and $r=-\infty$, respectively.  According to (\ref{phiform}), these functions must be related by
\begin{eqnarray}
\bar{\phi}_{+}(\vec{y}) &=& S \, \phi_+(\vec{y}), \label{BC1}\\
\bar{\phi}_{-}(\vec{y}) &=& -\frac{1}{S} \, \phi_{-}(\vec{y}) \label{BC2}
\end{eqnarray}
After substituting the above into (\ref{general1}) we obtain
\begin{equation}
\phi(r,\vec{y}) = \left\{ \begin{array}{cc} \int d^d \vec{y} \, ' \, K_{\Delta_+}(\vec{y},\vec{y} \, ',r) \, \phi_+(\vec{y} \, ')+\int d^d \vec{y} \, ' \, K_{\Delta_-}(\vec{y},\vec{y} \, ',r) \, \phi_-(\vec{y} \, '), & (r>0) \\
S \, \int d^d \vec{y} \, ' \, K_{\Delta_+}(\vec{y},\vec{y} \, ',-r) \, \phi_+(\vec{y} \, ')-\frac{1}{S} \, \int d^d \vec{y} \, ' \, K_{\Delta_-}(\vec{y},\vec{y} \, ',-r) \, \phi_-(\vec{y} \, '), & (r<0) \end{array} \right. \label{phiplus}
\end{equation}

Now the boundary conditions on $\phi$ are 
\begin{equation}
\phi(r,\vec{y}) = \left\{ \begin{array}{cc} f_+(\vec{y}) \, r^{-\Delta_-}+ f_-(\vec{y}) \, r^{-\Delta_+}+{\cal O}(r^{-2-\Delta_{\pm}})
 & (r \rightarrow +\infty), \\
 \bar{f}_+(\vec{y}) \, (-r)^{-\Delta_-}+  \bar{f}_-(\vec{y}) \, (-r)^{-\Delta_+}+{\cal O}(r^{-2-\Delta_{\pm}}) &  (r \rightarrow -\infty). \end{array} \right.  \label{phif}
\end{equation}
Here $f_{\pm}(\vec{y})$ and $\bar{f}_{\pm}(\vec{y})$ are functions which are determined in terms of $\phi_{\pm}(\vec{y})$. 
\begin{eqnarray}
f_+ &=& \phi_+(\vec{y})+\frac{\Gamma(\Delta_-)}{\pi^{\frac{d}{2}} \, \Gamma(-\nu)} \, \int \, |\vec{y}-\vec{y} \, '|^{-2\Delta_-} \, \phi_-(\vec{y} \, ') \, d^d \vec{y} \, ', \label{bcfp}\\
f_- &=& \phi_-(\vec{y})+\frac{\Gamma(\Delta_+)}{\pi^{\frac{d}{2}} \, \Gamma(\nu)} \, \int \, |\vec{y}-\vec{y} \, '|^{-2\Delta_+} \, \phi_+(\vec{y} \, ') \, d^d \vec{y} \, ',  
\end{eqnarray}
\begin{eqnarray}
\bar{f}_+ &=& S \, \phi_+(\vec{y})-\frac{1}{S} \, \frac{\Gamma(\Delta_-)}{\pi^{\frac{d}{2}} \, \Gamma(-\nu)} \, \int \, |\vec{y}-\vec{y} \, '|^{-2\Delta_-} \, \phi_-(\vec{y} \, ') \, d^d \vec{y} \, ', \\
\bar{f}_- &=& -\frac{1}{S} \, \phi_-(\vec{y})+S \,\frac{\Gamma(\Delta_+)}{\pi^{\frac{d}{2}} \, \Gamma(\nu)} \, \int \, |\vec{y}-\vec{y} \, '|^{-2\Delta_+} \, \phi_+(\vec{y} \, ') \, d^d \vec{y} \, '  \label{bcfm}
\end{eqnarray}
The first terms are source functions and the second terms are \lq responses' to the sources. 
In actual calculations of the asymptotics of a given solution, one cannot distinguish the two. 
For the integrals in $f_-$ and $\bar{f}_-$, some regularization for the singularities at $\vec{y}=\vec{y} \, '$ will be necessary. Now the ${\cal O}(r^{-\Delta_-})$ and ${\cal O}(r^{-\Delta_+})$ terms in $\phi(r,\vec{y})$ are fixed on the boundaries, and in the derivation of the equation of motion, the variation of $\phi$ is at most $\delta \phi(r,\vec{y})= {\cal O}(r^{-2-\Delta_-})$. Then the variations of the action on the boundaries vanish: $\int d^d \vec{y} \, \sqrt{g} \,  \phi \, r \, \partial_r  \, \delta \phi \rightarrow 0$, \\ $ \int d^d \vec{y} \, \sqrt{g} \,  \delta \, \phi \, r \, \partial_r  \,  \phi \rightarrow 0 \ (r \rightarrow \pm \infty)$. Hence the variational problem is well-posed. 
%By (\ref{bcfp}) - (\ref{bcfm}), $\phi_{\pm}$ are determined in terms of $f_{\pm}$ and $\bar{f}_{\pm}$: $(1+S^2) \, \phi_+=f_++S\bar{f}_+$ and $(1+S^{-2}) \phi_-=f_--S^{-1} \, \bar{f}_-$. There are two relations among $f_{\pm}$ and $\bar{f}_{\pm}$. 
To determine $f_{\pm}$ and $\bar{f}_{\pm}$ in terms of $\phi_{\pm}$, one needs to know $K_{\Delta_{\pm}}$.\footnote{ The mode functions $\psi_{\pm}$ (\ref{psi2}) have asymptotic behaviours $r^{-\frac{d}{2}\pm \nu}$, respectively. However, after integrating over the modes, each bulk-boundary propagator $K_{\Delta_{\pm}}$ acquires both power behaviours (\ref{Kasymp}).              
In order to impose the boundary condition on the scalar field, one needs to use $K_{\Delta_{\pm}}$.  }
In an asymptotically AdS space, such as the one in the presence of black holes, one would need to use a bulk-boundary propagator $K'_{\Delta_{\pm}}$ of a scalar field in such a background. 

%Note that this kind of complication also arises, even if one tries to quantize only one type of scalar modes corresponding to $\phi_+$. Asymptotic forms (\ref{phif}) contain both powers $r^{-\Delta_{\pm}}$. One cannot set $\phi_-=0$ by just specifying only $f_+$, since it depends on both $\phi_+$ and $\phi_-$, and ${\cal O}(r^{-\Delta_+})$ term $f_-$ is also non-vanishing. One needs to know the explicit forms of $K_{\Delta_{\pm}}$, and correlate $f_+$ and $f_-$.

Let us now turn to the behaviour of the solution (\ref{phiplus}) near the horizon. 
We consider  integrals $\int d^d \vec{y} \,' [r/(1+r^2(\vec{y}-\vec{y} \,')^2]^{\Delta_{\pm}} \, \phi_{\pm}(\vec{y} \,')$.  As far as the source functions $\phi_{\pm}(\vec{y})$ have compact supports, these integrals can be approximated as $r^{\Delta_{\pm}} \, \int d^d \vec{y} \,' \, \phi_{\pm}(\vec{y} \,')$ as $r \rightarrow 0$. 
Hence 
\begin{equation}
\int d^d \vec{y} \,' K_{\Delta_+}(\vec{y},\vec{y} \,',r) \, \phi_+(\vec{y} \,') \sim \frac{\Gamma(\Delta_{\pm})}{\pi^{\frac{d}{2}} \Gamma(\Delta_{\pm}-\frac{d}{2})} \Big[\int d^d \vec{y} \,' \phi_{\pm}(\vec{y} \,')\Big] \ r^{\Delta_{\pm}}
\end{equation}
and the solution $\phi$ to the boundary-value problem behaves near the boundary as $\phi \sim r^{\Delta_{\pm}}$, although the mode functions (\ref{modep}) and (\ref{moden}) are blowing up and oscillating rapidly near the horizon. Then, $\phi$ and $(r \, \partial_r)^n \, \phi=\partial_{\tilde{\rho}}^n \, \phi$ vanish on the horizon, and the surface terms on the horizon are not required.

\subsection{Two-point Functions and Boundary Terms}
\hspace{5mm}
According to AdS/CFT correspondence, in the semi-classical regime, the on-shell action of the scalar field in the AdS background is supposed to give generating functional of two-point functions of single-trace operators ${\it O}_+$ or ${\it O}_-$ in boundary CFT. In this paper we will try to realize  AdS/CFT correspondence for ${\it O}_+$ and ${\it O}_-$ altogether at the same time. 
Since the bulk action $I_0$ (\ref{bulkaction}) vanishes on shell, we need to introduce boundary terms (and counterterms). The choice of the boundary terms defines definite theories. 
Because there are two boundaries, we can introduce boundary action $I_{\pm}$ on each boundary. 
They must be local functionals of $\phi$ and its derivatives. We will consider the following form. 
\begin{eqnarray}
I_{\pm} &=& \pm \, \lim_{r \rightarrow \pm \infty}  \int_{r=\text{fixed}} d^d\vec{y} \, |r|^d \, \Big[ \alpha_1 \, \phi^2+ \alpha_2 \, \phi \, r \partial_r \, \phi+\alpha_3
\, ( r\partial_r \, \phi)^2\nonumber \\  &&\qquad \qquad \qquad +\alpha_4 \, \phi \,( r\partial_r)^2 \phi \big] \nonumber \\
&=& \pm \, \lim_{r \rightarrow \pm \infty}  \int_{r=\text{fixed}} d^d\vec{y} \, \Big[ \pm \alpha_2 \, \sqrt{g}\, \phi \, g^{rr} \, \partial_r \, \phi+ \sqrt{\gamma}\, \big\{\alpha_1 \, \phi^2+\alpha_3 \, g^{rr}
\, ( \partial_r \, \phi)^2\nonumber \\  && \qquad \qquad \qquad  +\alpha_4 \, g^{rr} \, \phi \, \partial_r^2 \phi\big\} \big] \label{boundaryS}
\end{eqnarray}
Here $\alpha_i$ ($i=1,2,3,4$) are constants, and $\gamma_{ij}$ is an induced metric on the boundaries. It will turn out that the generating functional is universal up to a multiplicative constant, and we can set $\alpha_3=\alpha_4=0$. 
These boundary terms are  invariant under reparametrizations which keep the boundary unchanged.

We will substitute the general solution (\ref{general1})-(\ref{general2})  into (\ref{boundaryS}). It is necessary to evaluate integrals of a form $\lim_{r \rightarrow \infty} \int d^d \vec{y} \, r^d \, (r\partial_r)^n \, K_{\Delta}(\vec{y},\vec{y}_1, r) \, (r\partial_r)^{n'} \, K_{\Delta'}(\vec{y},\vec{y}_2, r)$ with $n,n'=0,1,2$. The method will be explained in Appendix B.
Then, the on-shell boundary action $I_+$ (\ref{boundaryS}) is given by\footnote{We also evaluated these integrals using Fourier transforms of the bulk-boundary Green functions (\ref{bulkboundary}), $\tilde{K}_{\Delta}(\vec{k})=2^{\frac{d}{2}-\Delta+1}\, \Gamma(\Delta-\frac{d}{2})^{-1} \, r^{-\frac{d}{2}} \, |\vec{k}|^{\Delta-\frac{d}{2}} \, \text{K}_{\frac{d}{2}-\Delta}(|\vec{k}|/r)$, with identical results. $\text{K}_{d/2-\Delta}$ on the right hand side is a McDonald function.}
\begin{eqnarray}
I_+ &= & A_1 \, \int d^d \vec{y}_1 \,\int  d^d \vec{y}_2 \ \phi_+(\vec{y}_1) \, y_{12}^{-2\Delta_+} \, \phi_+(\vec{y}_2) \nonumber \\
&& +A_2 \, r^{2\nu}\int d^d \vec{y}_1 \,\int  d^d \vec{y}_2 \ \phi_+(\vec{y}_1) \, y_{12}^{-2\Delta_-} \, \phi_-(\vec{y}_2) \nonumber \\
&& +\int d^d \vec{y}_1 \,\int  d^d \vec{y}_2 \ \phi_-(\vec{y}_1) \, \Big(A_3 \, r^{2\nu} \, y_{12}^{d-4\Delta_-} +A_4 \, y_{12}^{-2\Delta_-} \Big)\, \phi_-(\vec{y}_2).
\end{eqnarray}
The coefficients $A_1, \cdots, A_4$ are given by
\begin{eqnarray}
A_1 &=& \frac{2\Gamma(\Delta_+)}{\pi^{\frac{d}{2}} \, \Gamma(\nu)} \, \Big[ \alpha_1-\frac{d}{2} \, \alpha_2  + \Delta_+ \, \Delta_- \, \alpha_3+\frac{1}{2} \, (\Delta_+^2+\Delta_-^2) \, \alpha_4 \Big],  \\
A_2 &=&  \frac{2\Gamma(\Delta_-)}{\pi^{\frac{d}{2}} \, \Gamma(-\nu)} \, \Big[  \alpha_1-\Delta_- \, \alpha_2+ \Delta_-^2 \, (\alpha_3+\alpha_4)\Big],  \\
A_3 &=& \frac{\Gamma(\frac{d}{2}-2\nu)\Gamma(\nu)^2}{\pi^{\frac{d}{2}} \, \Gamma(2\nu)\Gamma(-\nu)^2} \Big[  \alpha_1-\Delta_- \, \alpha_2+\Delta_-^2(\alpha_3+ \alpha_4) \Big], \\
A_4 &=&  \frac{2\Gamma(\Delta_-)}{\pi^{\frac{d}{2}} \, \Gamma(-\nu)} \, \Big[ \alpha_1-\frac{d}{2} \, \alpha_2+\Delta_+ \, \Delta_- \, \alpha_3+\frac{1}{2} \,  (\Delta_+^2 + \Delta_-^2) \, \alpha_4 \Big]. \label{As}
\end{eqnarray}
This result can also be obtained more easily by using (\ref{Kasymp}) and an integral formula 
\begin{multline}
\int d^d \vec{y} \ |\vec{y}-\vec{y}_1|^{-2\Delta} \, |\vec{y}-\vec{y}_2|^{-2\Delta'}   = y_{12}^{d-2\Delta-2\Delta'} \, \pi^{\frac{d}{2}} \, \frac{\Gamma(\Delta+\Delta'-\frac{d}{2}) \Gamma(\frac{d}{2}-\Delta) \Gamma(\frac{d}{2}-\Delta)}{\Gamma(\Delta)\Gamma(\Delta')\Gamma(d-\Delta-\Delta')}, 
\end{multline}
which is a result of analytic continuation. Especially, the following identities hold.
\begin{eqnarray}
&& \int d^d \vec{y} \, |\vec{y}-\vec{y}_1|^{-2 \Delta_+} \, |\vec{y}-\vec{y}_2|^{-2\Delta_-}= 0, 
\label{confpm} \\
&& \int d^d \vec{y} \, |\vec{y}-\vec{y}_1|^{-2 \Delta_-} \, |\vec{y}-\vec{y}_2|^{-2\Delta_-}=  \frac{\pi^{\frac{d}{2}} \,\Gamma(\nu)^2}{\Gamma(\Delta_-)^2 \Gamma(2\nu)} \, y_{12}^{-d+4\nu}  \label{confmm}
\end{eqnarray}

Those coefficients $A_2$ and $A_3$ in (\ref{As}), which multiplies those terms  divergent as $r \rightarrow \infty$, must vanish. 
These conditions put a constraint on  the parameters $\alpha_i$.  
\begin{equation}
\alpha_1-\Delta_- \, \alpha_2 +\Delta_-^2 \, (\alpha_3+ \alpha_4)=0.
\end{equation}
There are still free parameters in addition to an overall constant. 
Note that this finiteness prescription eliminates the coupling between $\phi_+$ and $\phi_-$. 

Finally we get
\begin{multline}
-I_+ =   \frac{4\nu}{\pi^{\frac{d}{2}}}  \, \Big(\alpha_2-2 \, \Delta_- \, \alpha_3-d \, \alpha_4\Big) \, 
 \int d^d \vec{y}_1 \, \int d^d \vec{y}_2  \\ 
 \, \Big[ \frac{\Gamma(\Delta_+)}{\Gamma(\nu)} \, y_{12}^{-2\Delta_+} \, \phi_+(\vec{y}_1) \, \phi_+(\vec{y}_2)+  \frac{\Gamma(\Delta_-)}{\Gamma(-\nu)} \, y_{12}^{-2\Delta_-} \, \phi_-(\vec{y}_1) \, \phi_- (\vec{y}_2) \Big].   \label{Iponshell}
\end{multline}
We believe that even if further boundary terms are introduced in (\ref{boundaryS}), the result for $-I_+$ is unique up to an overall constant.  
From (\ref{Iponshell}) we can read off the two-point functions $\langle {\it O}_{\pm} (y_1) \, {\it O}_{\pm} (y_2) \rangle$ in boundary CFT by means of functional differentiations of $-I_+$. This result shows that there is some kind of universality. Even if we add extra boundary terms to the action as in (\ref{boundaryS}), after suitable renormalization,  the result will be
 proportional to a universal generating function.   
There is, however,  a serious problem in the present case. Since $\Gamma(-\nu)$ is negative for $0 < \nu <1$, for any choice of $\alpha_2$, $\alpha_3$ and $\alpha_4$, $\langle {\it O}_{+} (\vec{y}_1) \, {\it O}_{+} (\vec{y}_2) \rangle$ or $\langle {\it O}_{-} (\vec{y}_1) \, {\it O}_{-} (\vec{y}_2) \rangle$ necessarily turns out negative. This would imply that CFT would be non-unitary. 

This problem is actually resolved, when we also use $I_-$ as given  in (\ref{boundaryS}) with an overall negative sign with respect to $I_+$. Due to the relative coefficients in (\ref{phiplus}), we have from (\ref{Iponshell}), 
\begin{multline}
-I_- = -  \frac{4\nu}{\pi^{\frac{d}{2}}}  \, \Big(\alpha_2-2 \, \Delta_- \, \alpha_3-d \, \alpha_4\Big) \, 
 \int d^d \vec{y}_1 \, \int d^d \vec{y}_2  \\ 
 \, \Big[ S^2 \, \frac{\Gamma(\Delta_+)}{\Gamma(\nu)} \, y_{12}^{-2\Delta_+} \, \phi_+(\vec{y}_1) \, \phi_+(\vec{y}_2)+  S^{-2} \, \frac{\Gamma(\Delta_-)}{\Gamma(-\nu)} \, y_{12}^{-2\Delta_-} \, \phi_-(\vec{y}_1) \, \phi_- (\vec{y}_2) \Big].  \label{Imonshell}
\end{multline}
The sum $I= I_++I_-$ yields the following two point functions.
\begin{eqnarray}
\langle {\it O}_{+} (\vec{y}_1) \, {\it O}_{+} (\vec{y}_2) \rangle &=& \frac{8\nu}{\pi^{\frac{d}{2}} }\, \frac{\Gamma(\Delta_+)}{\Gamma(\nu)}  \, \Big((\alpha_2-2 \, \Delta_- \, \alpha_3-d \, \alpha_4\Big) \,      \, (1-S^2) \, y_{12}^{-2\Delta_+},  \\
\langle {\it O}_{-} (\vec{y}_1) \, {\it O}_{-} (\vec{y}_2) \rangle &=& \frac{8 \nu}{\pi^{\frac{d}{2}} }\, \frac{\Gamma(\Delta_-)}{\Gamma(-\nu)}   \, \Big((\alpha_2-2 \, \Delta_- \, \alpha_3-d \, \alpha_4\Big) \,  \, \frac{(S^2-1)}{S^2} \, y_{12}^{-2\Delta_+}, \\
\langle {\it O}_{+} (\vec{y}_1) \, {\it O}_{-} (\vec{y}_2) \rangle &=& 0
\end{eqnarray}
To reinstate unitarity, we need to adjust the parameters such that $  \Big(\alpha_2-2\, \Delta_- \, \alpha_3-d \, \alpha_4\Big) \,  \, (1-S^2) >0$.  If the bulk action (\ref{bulkaction}) is rewritten into an ordinary symmetric form by partial integration, boundary terms $- \frac{1}{2} \, \int_{r \rightarrow +\infty} d^d \vec{y} \, |r|^d \, \phi \, r \partial_r \, \phi$ and  $-\frac{1}{2} \, \int_{r \rightarrow -\infty} d^d \vec{y} \, |r|^d \, \phi \, r \partial_r \, \phi$ will appear, and to cancel the first term we must set  $\alpha_2=\frac{1}{2}$. In this case the second term is not canceled. Then, $\alpha_3=\alpha_4=0$ and $\alpha_1=\frac{1}{2} \, \Delta_-$ will be the simplest choice of parameters. Hence, $S^2<1$. The above prescription is different from the previous ones.\cite{ads1}\cite{ads3}\cite{ads4}

To summarize, after partial integration, the action integral is given by 
%\begin{multline}
%I=  \int_{-\infty}^{\infty} dr \, \int d^d y \sqrt{g} \, \Big(-\frac{1}{2} \, \frac{1}{\sqrt{g}} \phi \, \partial_{\mu} (\sqrt{g} \, g^{\mu\nu} \, \partial_{\nu}\phi) +\frac{1}{2} \, m^2\phi^2\Big) \\
%+ \lim_{r \rightarrow + \infty}   \int_{r \ \text{fixed}} d^d\vec{y} \, r^d \, \frac{1}{2}  \, \big( \Delta_- \, \phi^2 +\phi \, r \, \partial_r \, \phi \big)\\
%-\lim_{r \rightarrow - \infty}   \int_{r \ \text{fixed}} d^d\vec{y} \, (-r)^d \, \frac{1}{2} \, \big( \Delta_- \, \phi^2 -\phi \, r \, \partial_r \, \phi \big).   
%\end{multline}
\begin{multline}
I=  \int_{-\infty}^{\infty} dr \, \int d^d y \sqrt{g} \, \Big(\frac{1}{2}  \,  g^{\mu\nu} \,\partial_{\mu} \phi\, \partial_{\nu}\phi+\frac{1}{2} \, m^2\phi^2\Big) \\
+ \lim_{r \rightarrow + \infty}   \int_{r \ \text{fixed}} d^d\vec{y} \, \sqrt{\gamma} \, \frac{1}{2}  \,  \Delta_- \, \phi^2  
-\lim_{r \rightarrow - \infty}   \int_{r \ \text{fixed}} d^d\vec{y} \, \sqrt{\gamma} \, \frac{1}{2} \,  \Delta_- \, \phi^2  \\
-\lim_{r \rightarrow - \infty}   \int_{r \ \text{fixed}} d^d\vec{y} \, \sqrt{\gamma} \, \phi \, r \, \partial_r \, \phi   \label{finalI}
\end{multline}

\section{Three-point functions}
\hspace{5mm}
The Euclidean Green function satisfies 
\begin{equation}
(\frac{1}{\sqrt{g}} \partial_{\mu} (\sqrt{g} g^{\mu\nu}\partial_{\nu})-m^2 ) \, G_E(x,x')=-\frac{1}{\sqrt{g}} \, \delta^{(d+1)}(x,x').
\end{equation}
Let us consider a $\lambda \, \phi^3$ interaction with $\lambda$ being of order of $1/N$.\cite{Martinec}
\begin{equation}
I_0 = \int^{\infty}_{-\infty} dr \int d^d \vec{y} \sqrt{g} \, \left( -\frac{1}{2} \phi \frac{1}{\sqrt{g}} \partial_{\mu}\, ( \sqrt{g} g^{\mu\nu}\partial_{\nu})\phi+\frac{1}{2} \, m^2 \, \phi^2+\frac{1}{3} \lambda \, \phi^3 \right)  
\label{Icubic}
\end{equation}
 Equation of motion $\Delta \, \phi-m^2 \, \phi=\lambda \, \phi^2$ can be solved by using the Green function $G_E$:
\begin{multline}
\phi(r,\vec{y}) = \int d^d \vec{y} \, ' K_{\Delta_+}(\vec{y}, \vec{y} \, ',r) \, \phi_+(\vec{y} \, ')+\int d^d \vec{y} \, ' K_{\Delta_-}(\vec{y},\vec{y} \, ',r) \, \phi_-(\vec{y} \, ') \\
-\lambda \, \int d^d\vec{y} \, ' \int^{\infty}_{-\infty} dr' \, \sqrt{g(r')} G_E(r,\vec{y};r',\vec{y} \, ') \, \phi(\vec{y} \, ',r')^2, \qquad (r>0),
\end{multline}
\begin{multline}
\phi(r,\vec{y}) = S \, \int d^d\vec{y} \, ' K_{\Delta_+}(\vec{y},\vec{y} \, ',-r) \, \phi_+(\vec{y} \, ')-\frac{1}{S} \, \int d^d\vec{y}\,' K_{\Delta_-}(\vec{y},\vec{y} \, ',-r) \, \phi_-(\vec{y} \, ') \\
-\lambda \, \int d^d \vec{y} \, ' \int^{\infty}_{-\infty} dr' \, \sqrt{g(r')} G_E(r,\vec{y};r',\vec{y}\, ') \, \phi(\vec{y} \, ',r')^2, \qquad (r<0).
\end{multline}
These equations can be solved by iterations.  Up to the first order in $\lambda$, the solution is given as follows.
\begin{multline}
\phi(r,\vec{y}) = \int d^d \vec{y} \, ' K_{\Delta_+}(\vec{y},\vec{y} \, ',r) \, \phi_+(\vec{y} \, ')+\int d^d\vec{y} \, ' K_{\Delta_-}(\vec{y},\vec{y} \, ',r) \, \phi_-(\vec{y} \, ') \\
-\lambda \,  \int^{\infty}_{0} dr' \, \int d^d\vec{y} \, ' \, \sqrt{g(r')} G_E(r,\vec{y};r',\vec{y} \, ') \, \Big[  \int d^d\vec{y} \, '' K_{\Delta_+}(\vec{y} \, ',\vec{y} \, '',r') \, \phi_+(\vec{y} \, '') \\   +\int d^d\vec{y} \, '' K_{\Delta_-}(\vec{y} \, ',\vec{y} \, '',r') \, \phi_-(\vec{y} \, '')    
\Big]^2 \\
-\lambda \,  \int^{\infty}_{0} dr' \, \int d^d\vec{y} \, ' \, \sqrt{g(r')} G_E(r,\vec{y};-r',\vec{y} \, ') \, \Big[  S \, \int d^d\vec{y} \, '' K_{\Delta_+}(\vec{y} \, ',\vec{y} \, '',r') \, \phi_+(\vec{y} \, '') \\   -\frac{1}{S} \, \int d^d\vec{y} \, '' K_{\Delta_-}(\vec{y} \, ',\vec{y} \, '',r') \, \phi_-(\vec{y} \, '')    
\Big]^2  +{\cal O}(\lambda^2), 
\qquad (r>0) \label{solp}
\end{multline}
\begin{multline}
\phi(r,\vec{y}) = S \, \int d^d\vec{y} \, ' K_{\Delta_+}(\vec{y},\vec{y} \, ',-r) \, \phi_+(\vec{y} \, ')-\frac{1}{S} \, \int d^d\vec{y} \, ' K_{\Delta_-}(\vec{y},\vec{y} \, ',-r) \, \phi_-(\vec{y} \, ') \\
-\lambda \,  \int^{\infty}_{0} dr' \, \int d^d\vec{y} \, ' \, \sqrt{g(r')} G_E(r,\vec{y};r',\vec{y} \, ') \, \Big[  \int d^d\vec{y} \, '' K_{\Delta_+}(\vec{y} \, ',\vec{y} \, '',r') \, \phi_+(\vec{y} \, '') \\   +\int d^d\vec{y} \, '' K_{\Delta_-}(\vec{y} \, ',\vec{y} \, '',r') \, \phi_-(\vec{y} \, '')    
\Big]^2 \\
-\lambda \,  \int^{\infty}_{0} dr' \, \int d^d\vec{y} \, ' \, \sqrt{g(r')} G_E(r,\vec{y};-r',\vec{y} \, ') \, \Big[  S \, \int d^d\vec{y} \, '' K_{\Delta_+}(\vec{y} \, ',\vec{y} \, '',r') \, \phi_+(\vec{y} \, '') \\   -\frac{1}{S} \, \int d^d\vec{y} \, '' K_{\Delta_-}(\vec{y} \, ',\vec{y} \, '',r') \, \phi_-(\vec{y} \, '')    
\Big]^2  +{\cal O}(\lambda^2). 
\qquad (r<0) \label{solm}
\end{multline}

\begin{figure}[htbp]
\begin{minipage}{0.5\hsize}
\begin{center}
\includegraphics[scale=0.4]{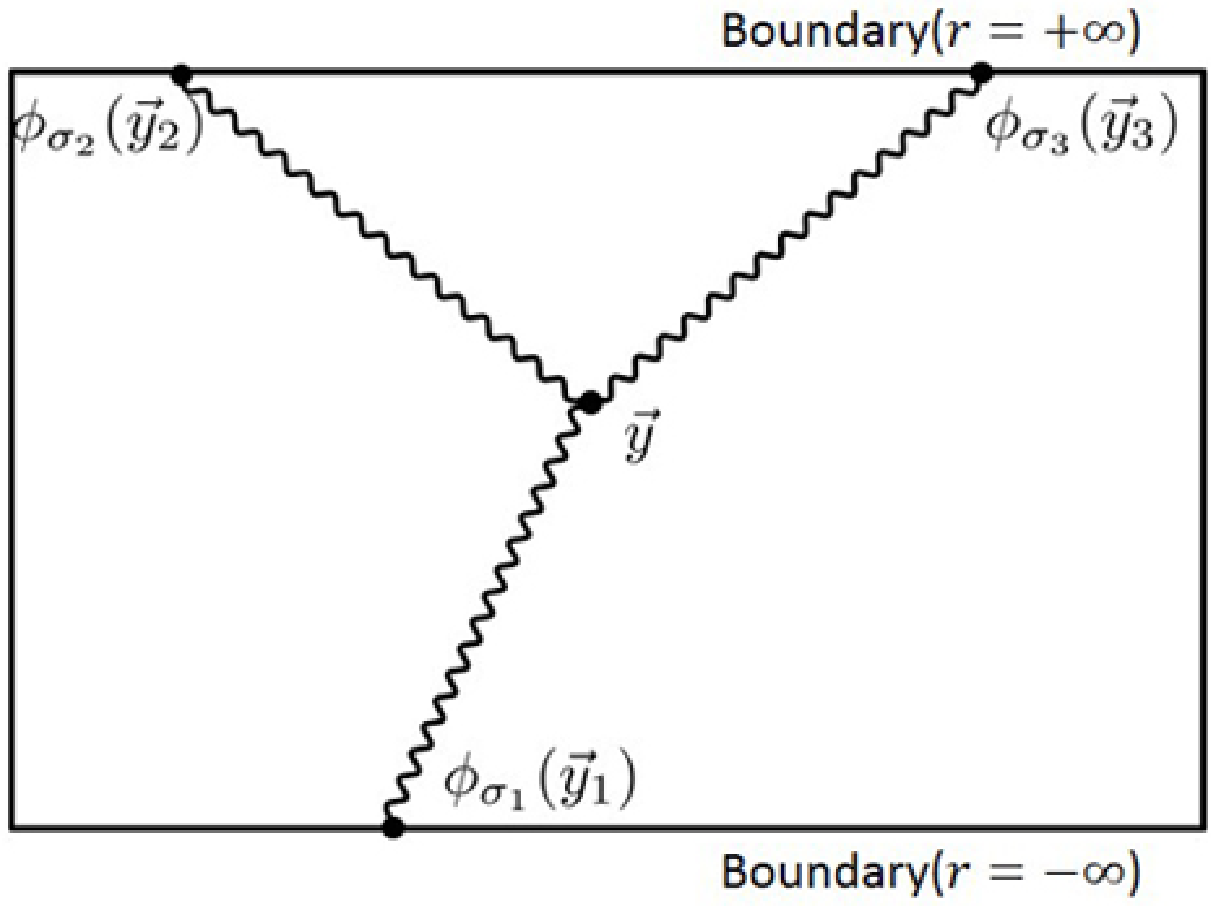} 
\end{center}
\caption{Graph contributing to three-point functions} \label{fig8}
\end{minipage}
\begin{minipage}{0.5\hsize}
\begin{center}
\includegraphics[scale=0.4]{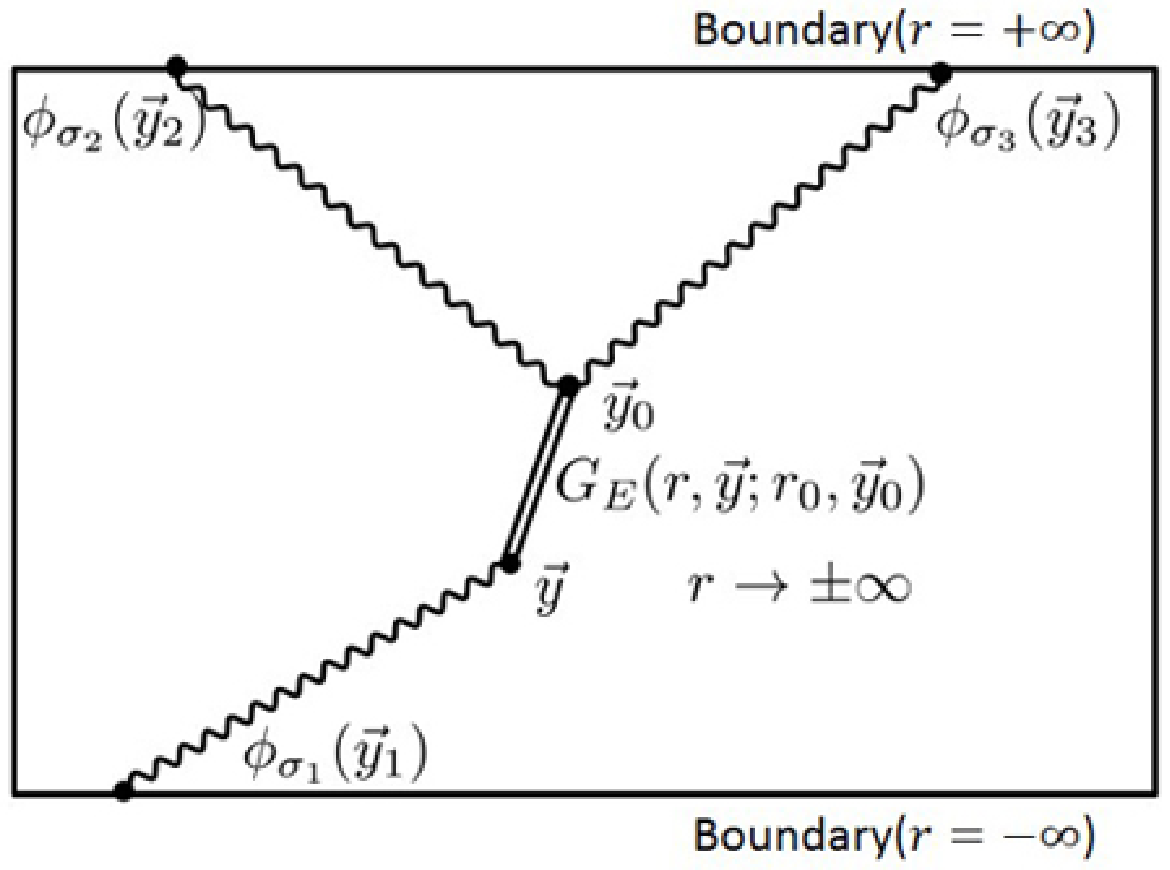} 
\end{center}
\caption{Graph contributing to three-point functions via boundary terms}  \label{fig9}
\end{minipage}
\end{figure}

By substituting equation of motion into the bulk part (\ref{Icubic}) we get a total action
\begin{multline}
I = \int_{-\infty}^{\infty} dr \, \int d^d\vec{y} \sqrt{g} \Big( -\frac{1}{6} \, \lambda \,  \phi^3 \Big) 
+ \int_{r \rightarrow +\infty} d^d \vec{y} \, r^d  \, (\frac{1}{2} \, \Delta_- \, \phi^2+\frac{1}{2} r \, \phi \, \partial_r \, \phi) \\
- \int_{r \rightarrow -\infty} d^d \vec{y} \, (-r)^d \, (\frac{1}{2} \, \Delta_- \, \phi^2+\frac{1}{2} r \, \phi \, 
\partial_r \, \phi).  \label{actioncube}
\end{multline}
The generating functional for three-point functions up to order ${\cal O}(\lambda^1)$ has two kinds of contributions: the bulk part and the boundary one. 

The bulk part is obtained by substituting the solution for the free theory (\ref{phiplus}) into the bulk action in (\ref{actioncube}). The corresponding diagram\cite{ads3} is presented in Figure \ref{fig8}. The wavy lines are bulk-boundary propagators $K_{\Delta_{\sigma}}$. There are actually a lot of terms and, a typical form of the terms is given by 
\begin{multline}
I^{(3, \text{bulk})} =\lim_{r \rightarrow \infty} \, \lambda \, \int_{0}^{\infty} dr \, r^{d-1} \int d^d \vec{y} \, \int d^d \vec{y}_1 \, \int d^d \vec{y}_2 \, \int d^d \vec{y}_3 \, K_{\Delta_1}(\vec{y},\vec{y}_1, |r|) \\ 
K_{\Delta_2}(\vec{y},\vec{y}_2, |r|) \, K_{\Delta_3}(\vec{y},\vec{y}_3, |r|) \, \phi_{\sigma_1}(\vec{y}_1) \, \phi_{\sigma_2}(\vec{y}_2) \, \phi_{\sigma_3}(\vec{y}_3) \label{bulk3point}
\end{multline}
Here $\sigma_i=\pm$ and $\Delta_i \equiv \Delta_{\sigma_i}$.  Integral of the form (\ref{bulk3point}) is evaluated in \cite{FMMR} by using an inversion, as 
\begin{multline}
I^{(3, \text{bulk})} =\lambda \, \int \,   \frac{a}{|\vec{y}_1-\vec{y}_2|^{\Delta_1+\Delta_2-\Delta_3} \,  |\vec{y}_1-\vec{y}_3|^{\Delta_1+\Delta_3-\Delta_2} \,  |\vec{y}_2-\vec{y}_3|^{\Delta_2+\Delta_3-\Delta_1} }  \\
 \phi_{\sigma_1}(\vec{y}_1) \, \phi_{\sigma_2}(\vec{y}_2) \, \phi_{\sigma_3}(\vec{y}_3) \, d^d\vec{y}_1 \, d^d\vec{y}_2 \, d^d\vec{y}_3,
\label{3bulk}
\end{multline}
where $a$ is a constant given by 
\begin{equation}
-\frac{\Gamma(\frac{1}{2}(\Delta_1+\Delta_2-\Delta_3)) \, \Gamma(\frac{1}{2}(\Delta_2+\Delta_3-\Delta_1)) \, \Gamma(\frac{1}{2}(\Delta_1+\Delta_3-\Delta_2))}{2\pi^d \, \Gamma(\Delta_1-\frac{d}{2}) \, \Gamma(\Delta_2-\frac{d}{2}) \, \Gamma(\Delta_3-\frac{d}{2}) } 
\Gamma(\frac{1}{2}(\Delta_1+\Delta_2+\Delta_3-d)).
\end{equation}

The boundary part of the generating function is obtained by substituting the solution (\ref{solp})-(\ref{solm}) into the boundary terms in (\ref{actioncube}). The corresponding diagram is depicted in Figure \ref{fig9}, and a typical form of the integrals is given by 
\begin{multline}
\lambda \, \lim_{r \rightarrow +\infty} \, \int d^d \vec{y} \, r^d \, \int d^d \vec{y}_1 \, \phi_{\sigma_1}(\vec{y}_1) \, \int_0^{\infty} dr_0 \, r_0^{d-1} \, \int d^d \vec{y}_0 \, F(\vec{y},\vec{y}_1,\vec{y}_0,r,r_0)  \\ \int d^d \vec{y}_2 \, K_{\sigma_2}(\vec{y}_0,\vec{y}_2,r_0) \, \phi_{\sigma_2}(\vec{y}_2) \, \int d^d \vec{y}_3 \, K_{\sigma_3}(\vec{y}_0,\vec{y}_3,r_0) \, \phi_{\sigma_3}(\vec{y}_3). \label{3boundary}
\end{multline}
Here $F$ is defined by
\begin{multline}
F(\vec{y},\vec{y}_1,\vec{y}_0,r,r_0) \equiv \Delta_- \, K_{\Delta_1}(\vec{y},\vec{y}_1,r) \, G_E(r,\vec{y};r_0,\vec{y}_0) \\ + \frac{1}{2} \, r\partial_r \, K_{\Delta_1}(\vec{y},\vec{y}_1,r) \, G_E(r,\vec{y};r_0,\vec{y}_0)+\frac{1}{2} \, K_{\Delta_1}(\vec{y},\vec{y}_1,r) \, r\partial_r \, G_E(r,\vec{y};r_0,\vec{y}_0).
\end{multline}
The propagator $G_E=G_++G_-$ with (\ref{GG1})-(\ref{GG2}) and (\ref{GK}) is to be substituted into (\ref{3boundary}). 
Upon substitution, each boundary term gives divergences. However, the linear combinations in the boundary terms work correctly, and the sum of all turns out finite. Moreover, the integral (\ref{3boundary}) can be explicitly carried out, and the result is proportional to the result of integral (\ref{3bulk}). The three-point functions are obtained by summing the bulk and boundary contributions. The details will be reported elsewhere. Here only the results of a three-point function of ${\it O}_+$ is presented. 
\begin{multline}
\langle {\it O}_+(\vec{y}_1) \, {\it O}_+(\vec{y}_2) \, {\it O}_+(\vec{y}_3) \rangle =\lambda \, 
\frac{-1+5 \, S^2-4 \, S^3+2 \, S^5}{1+S^2} \, \frac{\Gamma(\frac{\Delta_+}{2})^3 \, \Gamma\big(\frac{1}{2} \, (3\, \Delta_+-d)\big)}{\pi^{d/2} \, \Gamma(\nu)^3} \\
 \cdot \frac{1}{|\vec{y}_1-\vec{y}_2|^{\Delta_+} \,  |\vec{y}_1-\vec{y}_3|^{\Delta_+} \,  |\vec{y}_2-\vec{y}_3|^{\Delta_+} }
\end{multline}

\section{Discussion}
\hspace{5mm}
We showed a prescription for quantizing two sets of scalar modes in a pair of Poincar\'e patches of AdS space, and also presented a prescription for semi-classically obtaining  two- and three-point functions in the boundary CFT. This is possible since the two boundaries at $r=\pm \infty$ are connected, as a result of which the KG norm is conserved. 
Needless to say, more analysis is necessary. This will be left to future study. There are a few comments. 
 
If we want to quantize only a single set of scalar modes, or if $\nu >1$ and only modes $\hat{\psi}_+$ in  (\ref{psip}) are allowed, we can still do this in a pair of Poincar\'e patches. Mode expansion is (\ref{modephi}) with only $a_+$ and $a_+^{\dagger}$ retained. Canonical commutation relation is $[a_+(\omega,\vec{k}), a_+^{\dagger}(\omega',\vec{k}\,')]= 2^{-1-2\nu} \, (2\pi)^{-d+1} \, \Gamma(1+\nu)^{-2} \, \delta(\omega-\omega') \, \delta^{(d-1)}(\vec{k}-\vec{k}\,')$. Wightman function is proportional to $G_+$ in (\ref{Dpulus}), and the commutator $[\Phi(r,t,\vec{x}), \Pi(r',t,\vec{x}')]$ contains a term $iS \, \delta(r+r')\, \delta^{(d-1)}(\vec{x}-\vec{x}')$, which is harmless since a singularity at $r'=-r$ is beyond the horizon. 

If the two operators ${\it O}_+$ and ${\it O}_-$ are present in the boundary CFT, the sum of the scaling dimensions $\Delta_+$ and $\Delta_-$ is $d$, and by using a composite operator $\int d^d \vec{y} \, {\it O}_+{\it O}_-$, a marginal deformation of the CFT may be considered. A prescription for realizing this deformation in our formalism in the form of an interpolating geometry in $r$ direction is an interesting question. 

In the study of this paper, a parameter $S$ which parametrize quantization is introduced. The role of this parameter is to specify the relative magnitude of the mode functions in the two patches. We have not reached a concrete use of this degree of freedom yet. 
It is also discussed in sec.4 that by setting $S=0$ one can quantize only one of the two sets of the scalar modes on each of the two Poincar\'e patches.

The procedure of this paper can be extended to the black hole geometry. Schwarzschild-AdS$_{d+1}$ black hole solution in Poincar\'e coordinates is given by 
\begin{equation}
ds^2 = \ell^2 \Big(-f(r) \,  dt^2+\frac{1}{f(r)} \, dr^2+r^2 \, d\vec{x}^2\Big), \label{lineSAdS}
\end{equation}
where
\begin{equation}
f(r)= r^2 \, \big\{1-\big|\frac{r_+}{r}\big|^d \big\}.
\end{equation}
Event horizon is at $r=r_+(>0)$ and temperature is $T=\frac{d}{4\pi} \, r_+$. To quantize two sets of modes of a 
scalar field with mass $m$ in the range $ 0 <(m\ell)^2+\frac{d^2}{4} <1$ in this background, we consider a pair of Poincar\'e patches with $r>0$ and $r<0$.  Event horizons are at $r=\pm r_+$. The line element (\ref{lineSAdS}) is to be used in both patches. In Lorentzian space, the boundary conditions for scalar field at the event horizons must be in-going conditions. The fluxes across the horizons vanish due to $f(\pm r_+)=0$. At the boundaries, the boundary conditions for the scalar field must be such that the fluxes at the boundaries cancel out. These will be (\ref{BC1}) and (\ref{BC2}). 
It is interesting to compute partition functions and entropies for the black hole geometries. 

%\newpage
\setcounter{section}{0}
\renewcommand{\thesection}{\Alph{section}}

\section{Calculation of Wightman Function}
\hspace{5mm}
By substituting $\omega=\sqrt{\vec{k}^2+\mu^2}\equiv \omega(\vec{k}, \mu)$ into (\ref{Dpm}) with $r,r'>0$, we obtain
\begin{eqnarray}
G_+ &=& \frac{1}{(1+S^2) \, (rr')^{\frac{d}{2}}} \, \int_0^{\infty} d\mu \, \mu \,  J_{\nu}\Big(\frac{\mu}{r}\Big) \, J_{\nu}\Big(\frac{\mu}{r'}\Big) \, G^{\text{Flat}}(t,\vec{x}, t',\vec{x}';\mu). \label{Wight}
\end{eqnarray}
Here 
\begin{equation}
G^{\text{Flat}}(t,\vec{x}, t',\vec{x}';\mu)= \int \frac{d^{d-1} \, \vec{k}}{(2\pi)^{d-1}\, 2\omega(\vec{k},\mu)} \, e^{-i\omega(\vec{k},\mu)(t-t')+i\vec{k} \cdot (\vec{x}-\vec{x}')}  \label{FLAT}
\end{equation}
is a Wightman function for a free scalar field of mass $\mu$ in flat space.  
In the literature \cite{BFDG}, it was argued that the coincident-point singularity of Wightman function (or, Feynman propagator) in AdS space should agree with that in flat space and this fixes its normalization. By using the fact that Wightman function is a function of AdS-invariant distance and satisfies a certain differential equation, Wightman function was determined.  In this appendix, calculation of integral (\ref{Wight}) is explicitly carried out. 

For space-like separation of the plane coordinates, we can set $t-t'=0$ by using the Lorentz symmetry of the integral. We also set $\vec{x}\,'=\vec{0}$. 
It suffices to consider the case $r, r' >0$. 
To perform integration over $\vec{k}$, we note the following formulae.\cite{Grad} 
\begin{eqnarray}
\int_0^{\pi} e^{iz \cos\theta} \, \sin^{2\nu}\theta \, d\theta  &=& 
\sqrt{\pi} \, \Gamma(\nu+\frac{1}{2}) \, \Big(\frac{2}{z}\Big)^{\nu} J_{\nu}(z), \\
\int_0^{\infty} dx x^{\nu+1} \, (x^2+y^2)^{-\mu-1} \, J_{\nu}(ax) &=& 
\frac{a^{\mu} \, y^{\nu-\mu}}{2^{\mu} \Gamma(\mu+1)} \, \text{K}_{\nu-\mu}(ay), \nonumber \\
 && [\text{for } \ \ 2 \, \text{Re}(\mu)+\frac{3}{2}> \text{Re} (\nu)> -1]
\end{eqnarray}
$\text{K}_{\nu}(z)$ is McDonald function. By using these, we find that
\begin{equation}
\int \frac{d^{d-1}\vec{k}}{(2\pi)^{d-1}} \, \frac{1}{\sqrt{\vec{k}^2+\mu^2}} \, e^{i\vec{k} \cdot \vec{x}} 
= \frac{2}{(2\pi)^{\frac{d}{2}} }\, \Big(\frac{|\vec{x}|}{\mu}\Big)^{1-\frac{d}{2}} \, \text{K}_{\frac{d}{2}-1}(\mu \, |\vec{x}|). 
\end{equation}
This leads to 
\begin{multline}
G_+(r,\vec{x};r',\vec{0})  \\ = \frac{1}{(2\pi)^{\frac{d}{2} }\, (1+S^2) \, (rr')^{\frac{d}{2}}} \, \int_0^{\infty} d\mu \, \mu \,
\Big(\frac{|\vec{x}|}{\mu}\Big)^{1-\frac{d}{2}} \, \text{K}_{\frac{d}{2}-1}(\mu |\vec{x}|) \, J_{\nu}\Big(\frac{\mu}{r}\Big) \, J_{\nu}\Big(\frac{\mu}{r'}\Big). 
\end{multline}
Now by using the formulae \cite{Grad}
\begin{multline}
\int_0^{\infty}dx \, x^{\mu+1} \,  \text{K}_{\mu}(ax)J_{\nu}(bx)J_{\nu}(cx) \\=\frac{1}{\sqrt{2\pi}} \, a^{\mu} \, b^{-\mu-1} \, c^{-\mu-1} \, e^{-(\mu+\frac{1}{2})\pi i} (u^2-1)^{-\frac{1}{2}\, \mu-\frac{1}{4}} \, Q^{\mu+\frac{1}{2}}_{\nu-\frac{1}{2}}(u),  \\
[ 2bcu = a^2+b^2+c^2, \ \ \text{Re} (a) >|\text{Im} (b)|+|\text{Im}( c)|, \ \ \text{Re} \, (\nu )>-1, \ \ \text{Re}\,  (\mu+\nu) >-1], 
\end{multline}
\begin{equation}
Q^{\mu}_{\nu}(z)=\frac{e^{\mu \pi i} \, \Gamma(\nu+\mu+1) \, \Gamma(\frac{1}{2})}{2^{\nu+1} \, \Gamma(\nu+\frac{3}{2})} \, (z^2-1)^{\frac{\mu}{2}} \, z^{-\nu-\mu-1} \, _2F_1(\frac{\nu+\mu+2}{2},\frac{\nu+\mu+1}{2},\nu+\frac{3}{2};\frac{1}{z^2}),
\end{equation}
we obtain
\begin{multline}
G_+(r,\vec{x};r',\vec{0}) = \frac{\Gamma(\nu+\frac{d}{2})}{2(S^2+1)\pi^{\frac{d}{2}} \, \Gamma(\nu+1)} \, \frac{1}{(rr')^{\nu+\frac{d}{2}} \, (\frac{1}{r^2}+\frac{1}{r'^2}+\vec{x}^2)^{\nu+\frac{d}{2}}}  \\
 _2F_1(\frac{d}{4}+\frac{\nu+1}{2},\frac{d}{4}+\frac{\nu}{2},\nu+1;\frac{4}{(rr')^2} \, \frac{1}{
(\frac{1}{r^2}+\frac{1}{r'^2}+\vec{x}^2)^2 }).
\end{multline}

Finally by using a quadratic transform of a hypergeometric function,\cite{Grad} 
\begin{equation}
_2F_1(a,b,2b;2z)= (1-z)^{-a} \, _2F_1\Big(\frac{a}{2},\frac{a+1}{2},b+\frac{1}{2};\Big(\frac{z}{1-z}\Big)^2\Big), 
\end{equation}
we have
\begin{equation}
G_+(r,\vec{x};r',\vec{0}) = \frac{\Gamma(\nu+\frac{d}{2})}{2\pi^{\frac{d}{2}}(1+S^2)\Gamma(\nu+1)} \, P^{\nu+\frac{d}{2}} \, _2F_1(\frac{d}{2}+\nu,\nu+\frac{1}{2},2\nu+1;-4P).  \label{Gplus}
\end{equation}
$P$ is defined in (\ref{P}). Similar expression for $G_-$ can be obtained by replacement $\nu \rightarrow -\nu$ in $G_+$ and multiplying the result by $S^2$. 
For example, for $d=$ even, $G=G_++G_-$ behaves near the singularity $P \rightarrow \infty$ as 
\begin{equation}
G(r,\vec{x};r',\vec{0}) \rightarrow \frac{1}{4\pi^{\frac{d+1}{2}} }\, \Gamma \big(\frac{d-1}{2} \big) \, \Big[ rr' \, \Big\{\Big(\frac{1}{r}-\frac{1}{r'}\Big)^2+\vec{x}^2 \Big\}\Big]^{-\frac{d-1}{2}}.
\end{equation}
This does not depend on $S$, $\nu$ or $\Delta$. 
Hence normalization of the singularity of Wightman function cannot be used to fix the value of $S$. 
It can be checked that eq (\ref{Gplus}) with $S=0$ agrees with  eq (7.4) for  $i G_F(x,x')$ in \cite{BFDG} after replacements $\lambda_{\pm} = \Delta_{\pm}$, $n=d+1$, $a=\ell^{-1}=1$, $2/u=-4P$ and substitution $y^0-y' \, ^0=0$. 

\section{Calculation of the integrals necessary for evaluating boundary actions $I_{\pm}$}
\hspace{5mm}
Here the method for evaluating integrals of a form $\lim_{r \rightarrow \infty} \int d^d \vec{y} \, r^d \, (r\partial_r)^n \, K_{\Delta}(\vec{y},\vec{y}_1, r)$ $ (r\partial_r)^{n'} \, K_{\Delta'}(\vec{y},\vec{y}_2, r)$ will be explained for the cases $n=n'=0$. 
First, let us consider an integral,
\begin{equation}
L_1=\lim_{r \rightarrow \infty}\int d^d\vec{y} \, r^d \, K_{\Delta_+}(\vec{y},\vec{y}_1,r) \, K_{\Delta_+}(\vec{y},\vec{y}_2,r).
\end{equation}
We use Feynman's parameter-integral formula
\begin{equation}
\frac{1}{X_1^{m_1}} \cdots \frac{1}{X_n^{m_n}}= \int_0^1dt_1 \cdots \int_0^1 dt_n \, \delta(\sum_i t_i-1) \, \frac{\prod_i t_i^{m_i-1}}{[\sum_i t_i \, X_i ]^{\sum_i m_i} }\, \frac{\Gamma(\sum_i m_i)}{\prod_i \Gamma(m_i)}, \label{Fey}
\end{equation} 
and perform $y$ integration.\cite{FMMR}
\begin{equation}
L_1 = \frac{\Gamma(2\Delta_+-\frac{d}{2})}{\pi^{\frac{d}{2}} \, \Gamma(\Delta_+-\frac{d}{2})^2} \, r^{2\Delta_+} \, \int_0^1 dt \, \frac{t^{\Delta_+-1} (1-t)^{\Delta_+-1}}{[1+r^2t(1-t)y_{12}^2]^{2\Delta_+-\frac{d}{2}}}  \label{L1}
\end{equation}
Here $\vec{y}_{12}=\vec{y}_1-\vec{y}_2$. In the $r \rightarrow \infty$ limit,  regions near $t=0$ and $t=1$ have dominant contributions. These contributions from $0 \leq t \leq \epsilon$ and $1-\epsilon \leq t \leq 1$ with $\epsilon=r^{-1}$ can be evaluated by setting $t=r^{-2} y_{12}^{-2} z$ or $t=1-r^{-2} y_{12}^{-2} z$ and replacing $t$ integral by $z$ integral.  These two contributions have the same values and we have 
\begin{eqnarray}
&& 2 \times \frac{\Gamma(2\Delta_+-\frac{d}{2})}{\pi^{\frac{d}{2}} \, \Gamma(\Delta_+-\frac{d}{2})^2} \, r^{2\Delta_+} \, \int^{r \, y_{12}^2}_0 dz \, z^{\Delta_+-1} \, (r^{-2} y_{12}^{-2})^{\Delta_+} \frac{1}{(1+z)^{2\Delta_+-\frac{d}{2}}} \nonumber \\
& \rightarrow & 2 \, \frac{\Gamma(2\Delta_+-\frac{d}{2})}{\pi^{\frac{d}{2}} \, \Gamma(\Delta_+-\frac{d}{2})^2} \, B(\Delta_+,\Delta_+-\frac{d}{2}) \, y_{12}^{-2\Delta_+}.
\end{eqnarray}
Here $B(a,b)$ is Euler's beta function. There is also a region of $t$ which must be taken into account. 
For the region $\epsilon < t< 1-\epsilon$, we can replace $1+r^2 \, t(1-t) \, y_{12}^2$ in the denominator of (\ref{L1}) by $r^2 \, t(1-t) \, y_{12}^2$.  This yields a contribution proportional to 
\begin{equation}
r^{-2\nu} \, y_{12}^{d-4\Delta_+} \, \int^{1-\epsilon}_{\epsilon} dt \, [t(1-t)]^{-1-\nu}\sim
\frac{2}{\nu} \, r^{-\nu} \, y_{12}^{d-4\Delta_+}
\end{equation}
For $\nu >0$ this damps in the $r \rightarrow \infty$ limit. 
So finally, we obtain a finite result. 
\begin{equation}
L_1 = \frac{2\, \Gamma(\Delta_+)}{\pi^{\frac{d}{2}}\, \Gamma(\Delta_+-\frac{d}{2})}  \, \, y_{12}^{-2\Delta_+}
\end{equation}
In the above calculation it is assumed that $\vec{y}_1 \neq \vec{y}_2$, and ultra local terms such as $\delta^{(d)} (\vec{y}_1-\vec{y}_2)$ are neglected. 

We then consider an integral
\begin{eqnarray}
L_2&=& \lim_{r \rightarrow \infty} \,  \int d^d\vec{y} \, r^d \, K_{\Delta_+}(\vec{y},\vec{y}_1,r) \, K_{\Delta_-}(\vec{y},\vec{y}_2,r) \nonumber \\ 
&=& \frac{\Gamma(\frac{d}{2})}{\pi^{\frac{d}{2}} \, \Gamma(\Delta_+-\frac{d}{2})\Gamma(\Delta_--\frac{d}{2})} \, \int_0^1 dt \, \frac{r^{d} \, t^{\Delta_+-1} (1-t)^{\Delta_--1}}{[1+r^2t(1-t)y_{12}^2]^{\frac{d}{2}}}  \label{L2}
\end{eqnarray}
Contribution from the region  $0 \leq t \leq \epsilon$ to the integral is 
\begin{eqnarray}
r^{-2\nu} y_{12}^{-2\Delta_+} \, \int_0^{ry_{12}^2} dz \, z^{\Delta_+-1} \, (1+z)^{-\frac{d}{2}}
\sim r^{-2\nu} \, y_{12}^{-2\Delta_+} \frac{1}{\nu}(ry_{12}^2)^{\nu} \rightarrow 0
\end{eqnarray}
The one from the region  $1-\epsilon \leq t \leq 1$ is 
\begin{equation}
r^{2\nu} \, y_{12}^{-2\Delta_-} \int_0^{ry_{12}^2} dz z^{\Delta_--1} (1+z)^{-\frac{d}{2}}.
\end{equation}
Although the integral is finite in the limit $r \rightarrow \infty$,  we need to take ${\it O}(r^{-\nu})$ correction into account because of the prefactor $r^{2\nu}$: 
\begin{equation}
 r^{2\nu} \, y_{12}^{-2\Delta_-} \, \Big[B(\Delta_-,\frac{d}{2}-\Delta_-)-\frac{1}{\nu} r^{-\nu}y_{12}^{-2\nu}+{\it O}(r^{-\nu-1} )\Big]
\end{equation}
From the region $\epsilon < t< 1-\epsilon$, we obtain contribution 
\begin{equation}
\int_{\epsilon}^{1-\epsilon}dt \, y_{12}^{-d} \, t^{\nu-1}(1-t)^{-\nu-1} \sim \frac{1}{\nu}y_{12}^{-d} r^{\nu}
\end{equation}
Hence we get 
\begin{equation}
L_2= \frac{\Gamma(\Delta_-)}{\pi^{\frac{d}{2}} \, \Gamma(\Delta_--\frac{d}{2})} \, y_{12}^{-2\Delta_-} \, r^{2\nu}.
\end{equation}

A final example is 
\begin{eqnarray}
L_3&=& \lim_{r \rightarrow \infty} \, \int d^d\vec{y} \, r^d \, K_{\Delta_-}(\vec{y},\vec{y}_1,r) \, K_{\Delta_-}(\vec{y},\vec{y}_2,r) \nonumber \\ 
&=& \frac{\Gamma(2\Delta_--\frac{d}{2})}{\pi^{\frac{d}{2}} \, \Gamma(\Delta_--\frac{d}{2})^2}  \, \int_0^1 dt \, \frac{r^d \, t^{\Delta_--1} (1-t)^{\Delta_--1}}{[1+r^2t(1-t)y_{12}^2]^{2\Delta_--\frac{d}{2}}}.  \label{L3}
\end{eqnarray}
Actually, $\vec{y}$ integration converges only for $d-4\Delta_- <0$. If $d \geq 4$, this condition is satisfied, because $0 < \nu <1$. Otherwise, $\nu$ must be in the range $0< \nu < d/4$. At the end of the calculation, we will analytically continue the results in variable $\nu$ to its remaining region.
Contribution to the integral from regions $0 \leq t \leq \epsilon$ and $1-\epsilon \leq t \leq 1$ is 
\begin{eqnarray}
2 \, y_{12}^{-2\Delta_-} \, \int_0^{ry_{12}^2} \, dz \, z^{\Delta_--1} \, (1+z)^{\frac{d}{2}-2\Delta_-}.
\end{eqnarray}
This is divergent as $r \rightarrow \infty$ and  is expanded as  
\begin{eqnarray}
2 \, y_{12}^{-2\Delta_-} \Big[  \frac{1}{\nu} r^{\nu} y_{12}^{2\nu}+B(\Delta_-,-\nu) \Big].
\end{eqnarray}
From integral in the region $\epsilon < t< 1-\epsilon$, we obtain 
\begin{equation}
r^{2\nu} \, y_{12}^{d-4\Delta_-} \, \int_{\epsilon}^{1-\epsilon} dt \Big[t(1-t)\Big]^{\nu-1}.
\end{equation}
This is expanded as 
\begin{eqnarray}
 r^{2\nu} \, y_{12}^{d-4\Delta_-} \ \Big\{B(\frac{d}{2}-\Delta_-,\frac{d}{2}-\Delta_-)-\frac{2}{\nu}  r^{-\nu}\Big\}.
\end{eqnarray}
Hence we get 
\begin{equation}
L_3= \frac{\Gamma(2\Delta_--\frac{d}{2})\Gamma(\nu)^2}{\pi^{\frac{d}{2}} \, \Gamma(2\nu) \,\Gamma(-\nu)^2 } \, r^{2\nu} \, y_{12}^{4\Delta_--d} +2\, \frac{\Gamma(\Delta_-)}{\pi^{\frac{d}{2}} \, \Gamma(-\nu)} \, y_{12}^{-2\Delta_-}. 
\end{equation}
In a similar way cases $n, n' \neq 0$ can also be worked out.

\end{document}